\documentclass{aa}  
\pdfoutput=1
\usepackage{graphicx}
\usepackage{txfonts}
\usepackage{amsmath} 
\usepackage{amsfonts}
\usepackage {amssymb}
\usepackage{multirow}
\usepackage{float}

\bibpunct{(}{)}{;}{a}{}{,}

\begin{document} 

   \title{Constraining galaxy cluster temperatures and redshifts with \textit{eROSITA} survey data}


   \author{K. Borm
          \inst{1}
          \and T. H. Reiprich\inst{1}
          \and I. Mohammed\inst{2}
          \and
          L. Lovisari\inst{1}
          }

   \institute{Argelander Institute for Astronomy, University of Bonn,
             Auf dem H\"ugel 71, 53121 Bonn, Germany
             \and
              Institute for Theoretical Physics, University of Z{\"u}rich,
              Winterthurerstrasse 190, 8057 Z{\"u}rich, Switzerland\\
              \\
              \email{kborm@astro.uni-bonn.de
              }
             }

   \date{Accepted April 15, 2014}

	\abstract
   {The nature of dark energy is imprinted in the large-scale structure of the Universe and thus in the mass and redshift distribution of galaxy clusters. The upcoming \textit{eROSITA} mission will exploit this method of probing dark energy by detecting $\sim100,000$ clusters of galaxies in X-rays.}
   {For a precise cosmological analysis the various galaxy cluster properties need to be measured with high precision and accuracy. To predict these characteristics of \textit{eROSITA} galaxy clusters and to optimise optical follow-up observations, we estimate the precision and the accuracy with which \textit{eROSITA} will be able to determine galaxy cluster temperatures and redshifts from X-ray spectra. Additionally, we present the total number of clusters for which these two properties will be available from the \textit{eROSITA} survey directly.}
   {We simulate the spectra of galaxy clusters for a variety of different cluster masses and redshifts while taking into account the X-ray background as well as the instrumental response. An emission model is then fit to these spectra to recover the cluster temperature and redshift. The number of clusters with precise properties is then based on the convolution of the above fit results with the galaxy cluster mass function and an assumed \textit{eROSITA} selection function.}
   {During its four years of all-sky surveys, \textit{eROSITA} will determine cluster temperatures with relative uncertainties of $\Delta T/T\lesssim10\%$ at the $68\%$-confidence level for clusters up to redshifts of $z\sim0.16$ which corresponds to $\sim1,670$ new clusters with precise properties. Redshift information itself will become available with a precision of $\Delta z/(1+z)\lesssim10\%$ for clusters up to $z\sim0.45$. Additionally, we estimate how the number of clusters with precise properties increases with a deepening of the exposure.\\ For the above clusters, the fraction of catastrophic failures in the fit is below $20\%$ and in most cases it is even much smaller. Furthermore, the biases in the best-fit temperatures as well as in the estimated uncertainties are quantified and shown to be negligible in the relevant parameter range in general. For the remaining parameter sets, we provide correction functions and factors. In particular, the standard way of estimating parameter uncertainties significantly underestimates the true uncertainty, if the redshift information is not available.
   }
   {The \textit{eROSITA} survey will increase the number of galaxy clusters with precise temperature measurements by a factor of $5-10$. Thus the instrument presents itself as a powerful tool for the determination of tight constraints on the cosmological parameters. At the same time, this sample of clusters will extend our understanding of cluster physics, e.g. through precise $L_{\text{X}}-T$ scaling relations.}

	\keywords{X-rays: galaxies: clusters --
                Galaxies: clusters: spectral analysis --
                large-scale structure of Universe --
                Methods: statistical
               }

	\maketitle


\section{Introduction}	

	Over the past years, galaxy clusters have become reliable cosmological probes for studying dark energy and for mapping the large-scale structure (LSS) of the Universe \citep[e.g.,][]{Borgani2001,Voit2004,Vikhlinin2009, Vikhlinin2009I, Mantz2010, Allen2011}. Further improved constraints on the nature of dark energy require the analysis of a large sample of galaxy clusters with precisely and accurately known properties. The future \textit{eROSITA} (\textbf{e}xtended \textbf{Ro}entgen \textbf{S}urvey with an \textbf{I}maging \textbf{T}elescope \textbf{A}rray) telescope \citep{Predehl2010,Merloni2012}, which is scheduled for launch in late 2015, will provide such a data sample \citep{Pillepich2011}.\\
\indent
X-ray observations of galaxy clusters allow for the precise determination of various cluster properties such as e.g. the total mass as well as the gas mass of the cluster or the temperature and the metal abundance of the intra-cluster medium (ICM) \citep[e.g.][]{Henriksen1986,Sarazin1986,Vikhlinin2009}. The information on these properties is imprinted in the emission spectrum of the ICM, which follows a thermal bremsstrahlung spectrum superimposed by emission lines of highly ionised metals \citep[e.g.,][]{Sarazin1986}. Especially notable are the Fe-L and the Fe-K line complexes at energies of $\sim1$ keV and $\sim7$ keV, respectively. For low gas temperatures of k$T\lesssim 2.5$ keV, emission lines are prominent features in the spectrum in the energy range of roughly ($0.5-8$) keV. With increasing temperatures the lines at the lower energies fade as the metals become completely ionised, whereas other emission lines, such as e.g. the hydrogen like Fe-K line, increase with higher gas temperatures \citep[e.g., Fig. 2 in][]{Reiprich2013}. Analogously to the temperature, the spectrum also reflects the density and the metallicity of the ICM, as well as the cluster redshift, which allows these properties to be recovered in the analysis of X-ray data. While very precise redshifts with uncertainties of $\Delta z \ll0.01$ can be obtained in optical spectroscopic observations, estimating redshifts from X-ray data directly allows for an optimisation of these time-consuming optical spectroscopic observations.\\
\indent
Cosmological studies based on galaxy clusters are especially dependent upon the information on their redshift and total mass. As the cluster mass is not a direct observable, galaxy cluster scaling relations are commonly applied to estimate this property based on e.g., the ICM temperature and the cluster redshift \citep[e.g.,][]{Vikhlinin2009,Pratt2009,Mantz2010,Reichert2011,Giodini2013}. This then allows for an analysis of the distribution of galaxy clusters with mass and redshift. This galaxy cluster mass function traces the evolution of the large-scale structure (LSS) and is highly dependent on the cosmological model, implementing galaxy clusters as cosmological probes \citep[e.g.,][]{Press1974,Tinker2008}. Testing the cosmological model through the study of the galaxy cluster mass function has become an important method within the past years \citep[e.g.,][]{Reiprich2002,Voit2004,Vikhlinin2009,Vikhlinin2009I,Mantz2010}. This analysis methodology is not only based on X-ray obervations, but can as well be applied to \textit{Sunyaev-Zel'dovich} (SZ) observations of galaxy clusters. Current SZ cluster surveys, performed by e.g. the \textit{Atacama Cosmology Telescope} (ACT), the \textit{South Pole Telescope} (SPT) and \textit{Planck}, are increasing the impact of these observations and already lead to an improvement in constraining the cosmological parameters \citep[e.g.][]{Vanderlinde2010,Planck2013,Reichardt2013}. Additionally, a combination of SZ ans X-ray observations allows for the calibration of hydrostatic cluster masses, which in turn improves the cosmological constraints. The \textit{eROSITA} instrument will soon improve the data sample of available X-ray clusters, in terms of precision, accuracy and number of clusters. This sample will thus especially allow for optimated cosmological studies by means of X-ray galaxy clusters. As a side effect, also future SZ observations will profit from this cluster sample.\\
\indent
\textit{eROSITA} is the German core instrument aboard the Russian Spektrum-Roentgen-Gamma (SRG) satellite, which is scheduled for launch in late 2015 \citep{Predehl2010,Merloni2012}. The main science driver of this mission is studying the nature of dark energy. The first four years of the mission are dedicated to an all-sky survey, followed by a pointed observation phase, both in the X-ray energy range between $(0.1-10)$ keV. Within the all-sky survey, a conservatively estimated effective average exposure time of $t_{\text{exp}}=1.6$ ks is achieved and we expect to detect a total of $\sim10^5$ galaxy clusters, including basically all massive clusters in the observable universe with $M\gtrsim3\times10^{14}h^{-1}\text{ M}_{\odot}$ \citep{Pillepich2011}. For these calculations a minimum of $50$ photon counts within the energy range of $0.5-2.0$ keV is assumed for the detection of a cluster. With this predicted data sample, current simulations estimate an increased precision of the dark energy parameters to $\Delta w_{\text{0}}\approx0.03$ (for $w_{\text{a}}=0$) and $\Delta w_{\text{a}}\approx0.20$ \citep[][; Pillepich et al., in prep.]{Merloni2012}, assuming an evolution of the equation of state of dark energy with redshift as $w_{\text{DE}}=w_{\text{0}}+w_{\text{a}}/(1+z)$.\\\indent
These forecasts consider only the galaxy cluster luminosity and redshift to be known with an assumed uncertainty, whereas the precision on the cosmological parameters will be improved if additional cluster information, such as e.g. the ICM temperature, is available \citep[compare e.g.,][]{Clerc2012}. In this work we thus present how accurately and precisely \textit{eROSITA} will be able to determine the ICM temperature in dependence on the cluster masses and redshifts. In an analogous simulation, we investigate for which clusters the survey data will allow for a redshift estimate to optimise optical follow-up observations \citep[compare e.g.,][]{Yu2011}. \\\indent
The outline of this paper is as follows. In section 2, we define the properties of the clusters included in our simulations. We also introduce the applied model for the X-ray background as well as the simulation and analysis methods. The following section presents the predicted precisions and accuracies for the cluster temperatures and redshifts, while section 4 emphasises on the number of clusters for which precise properties will be available from \textit{eROSITA} data. The final two sections 5 and 6 contain the discussion and conclusion of this work, respectively.\\\indent
If not stated otherwise, we apply a fiducial cosmology of $H_{ \text{0}}=100\cdot h$ km/s/Mpc with $h=0.7$, $\Omega_{\text{m}}=0.3$, $\Omega_{\Lambda}=0.7$, $\sigma_{ \text{8}}=0.795$ and the solar metallicity tables by \cite{Anders1989}.


\section{Simulation method and analysis}

The predictions for the cluster temperatures and redshifts are based on the analysis of galaxy cluster spectra for which we apply the software \textbf{\textit{xspec}} \citep{Arnaud1996} version 12.7.0. For the simulation of the spectra the cluster temperature, luminosity, redshift, metallicity and the foreground absorption need to be known as well as the background emission observed by \textit{eROSITA} and the instrumental response (RSP) of the detector. The RSP applied in our simulations contains the combined resolution of all seven telescopes averaged over the entire field-of-view.

\subsection{Cluster properties}
\label{ClusterProp}

For the clusters included in our simulations, we define the total mass $M_{500}$ and the redshift $z$ within the ranges of $13\leqslant\log(M/\text{M}_{\odot})\leqslant15.7$ and $-2\leqslant\log(z)\leqslant0.25$ in logarithmic steps of 0.15, which is equivalent to $10^{13}\leqslant M/\text{M}_{\odot}\leqslant5\times10^{15}$ and $0.01\leqslant z\leqslant1.78$, respectively. Based on these two input parameters, the remaining cluster properties are estimated through galaxy cluster scaling relations, where we apply the findings by \cite{Reichert2011}
\begin{eqnarray}
T\quad[\text{keV}]&=&\left(\frac{M}{10^{14}\text{ M}_{\odot}}\cdot 3.44\right)^{0.62}\cdot E(z)^{0.64} \label{EqT-M}\\
L_{\text{X}}\quad[10^{44}\text{erg/s}]&=& \left(\frac{M}{10^{14}\text{ M}_{\odot}}\cdot 0.61\right)^{1.92}\cdot E(z)^{1.7}\quad ,\label{EqL-M}
\end{eqnarray}
with the bolometric luminosity $L_{\text{X}}$ measured in the energy range between ($0.01-100$) keV and the redshift evolution
\begin{equation}
E(z)=[\Omega_{\text{m}}(1+z)^{3}+\Omega_{\Lambda}]^{1/2}\quad .
\end{equation}
This scaling relation presents itself as most conservative approach for high redshift clusters when compared to other works, e.g. \cite{Vikhlinin2009} and \cite{Pratt2009} (see Sec. \ref{Discussion}). Note that we neglect the intrinsic scatter in the scaling relations for our simulations to only focus on the performance of the instrument. However, for the computation of cosmological parameters by means of galaxy cluster data, this intrinsic scatter needs to be taken into account.\\
\begin{figure}
	\centering
	\includegraphics[scale=0.36,trim=0.0cm 0.4cm 0.0cm 0.8cm, clip=true ]{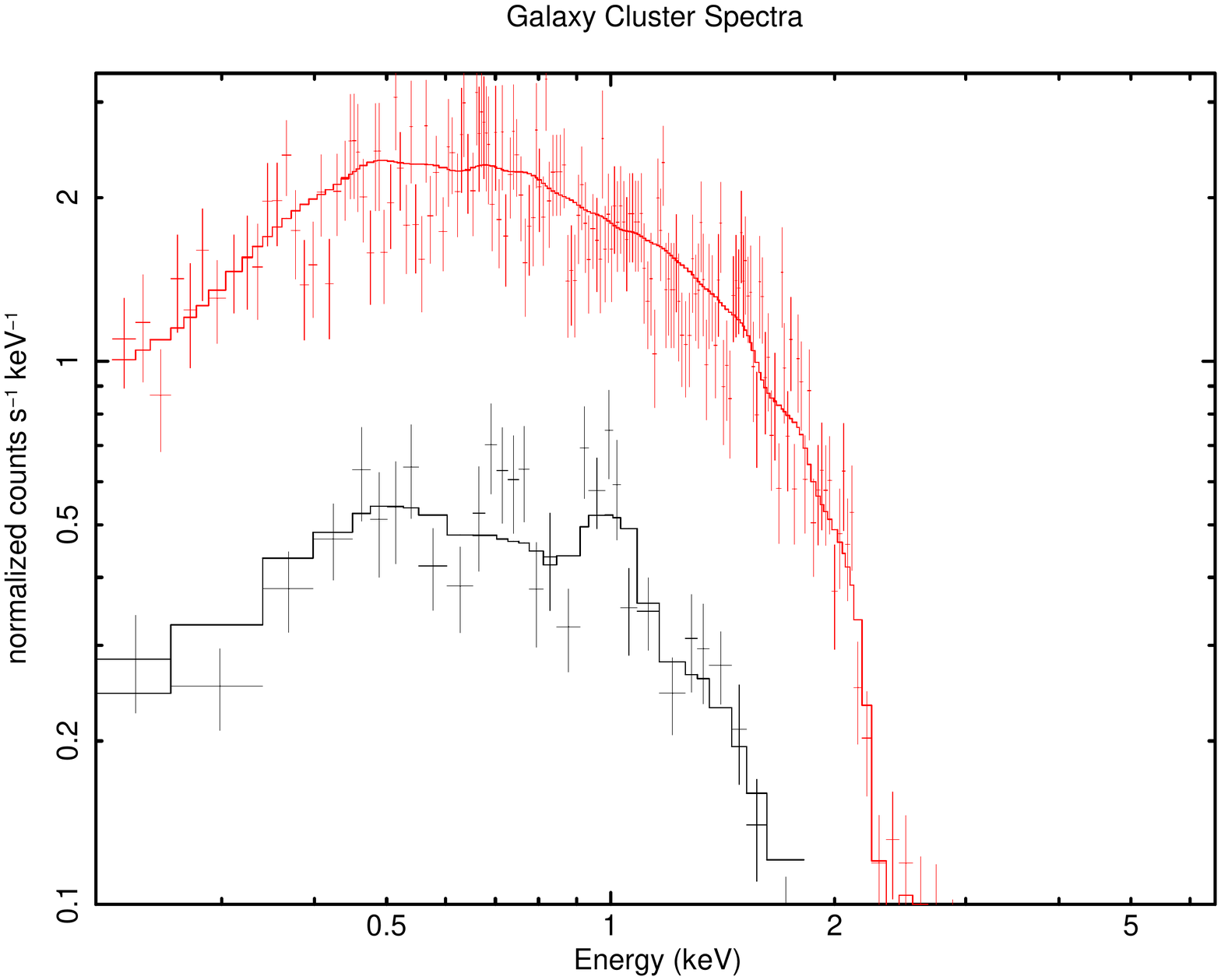}
	\caption{Simulated spectra of a galaxy cluster with $M_{500}=10^{14}\text{ M}_{\odot}$, $\text{k}T=2.2$ keV, and with $M_{500}=10^{15}\text{ M}_{\odot}$, $\text{k}T=9.8$ keV, respectively. The spectra are simulated for a redshift of $z=0.1$ and $z=0.3$, respectively, and for an exposure time of $t_{\text{exp}}=1.6$ ks. The model emission convolved with the instrumental response (continuous line) as well as the simulated emission (data points) are presented. For the simulated emission the energy bins are regrouped to yield at least 20 photons per group for display reasons.}
	\label{PixClusterSpectrum}
\end{figure}
Throughout all simulations, the cluster metallicity is set to $A=0.3\text{ A}_{\odot}$, which is a commonly observed value for nearby clusters \citep[e.g.][]{Arnaud1992,Mushotzky1997}.Though an evolution of the metallicity with redshift was observed, it could not be definitely quantified, yet \citep{Balestra2007,Maughan2008,Baldi2012}, and we thus prefer to apply the constant metallicity stated above. For a more detailed treatment of this evolution, we include a discussion of the effect of an abundance evolution with redshift in section \ref{FurtherUncertain}. At the same time, we assume the absorbing column density to be $N_{\text{H}}=3\times10^{20}$ particles/cm$^2$ as typical value for regions at galactic latitudes of $b\gtrsim20^{\circ}$ \citep{Kalberla2005}, which are relevant for the \textit{eROSITA} cluster survey. Figure \ref{PixClusterSpectrum} presents two example galaxy cluster spectra simulated as an absorbed thermal emission \citep{Smith2001} \textbf{phabs*apec}, convolved with the \textit{eROSITA} response. All clusters are simulated to show an isothermal emission.\\\indent
Furthermore, the simulations focus on clusters with fluxes below the \textit{eHIFLUGCS} limit of $9\times10^{-12}$ erg/s within the energy range of ($0.1-2.4$) keV (Schellenberger et al., in prep.). All clusters in this complete all-sky sample have high quality \textit{Chandra} and/or \textit{XMM} observations and, therefore, temperatures and redshifts are known. For clusters below this flux lumit no precise and accurate properties are usually available. At the same time, only clusters with a minimum of $100$ detected photons by \textit{eROSITA} in the energy range of ($0.3-8$) keV are considered to ensure a stable performance of the applied software. What is more, no reliable temperature and redshift measurements are expected for clusters with this low number of source events.\\
Even though the angular extension of the cluster does not define its over-all spectral emission, the extension is essential for the simulation of the X-ray background as the background normalisation is proportional to the observed region. The angular extension of the galaxy cluster is determined as $\alpha_{500}$ in dependence on the cluster mass and redshift
\begin{eqnarray}
	M_{500}&=&\frac{4\pi}{3}\rho_{\text{crit}}(z)\cdot500\cdot R_{500}^3\\
	\alpha_{500}&=&\frac{R_{500}}{D_{\text{A}}(z)}\quad,
\end{eqnarray}
applying the critical density $\rho_{\text{crit}}$ and the angular diameter distance $D_{\text{A}}$
\begin{eqnarray}
	\rho_{\text{crit}}&=&\frac{3H(z)^2}{8\pi \text{G}}\quad\text{with}\quad H(z)^2 = H_{\text{0}}\cdot E(z)^2\\
	D_{\text{A}}(z)&=&\frac{c}{H_{\text{0}}(1+z)}\int_0^zE(z)^{-1}\quad\text{d}z\quad.
\end{eqnarray}

\subsection{The \textit{eROSITA} X-ray background}
\label{background}
	
The background, observed by \textit{eROSITA}, is simulated following the modelled emission\\
\begin{equation}
\nonumber
\underbrace{\textbf{phabs}}_1*(\underbrace{\textbf{powerlaw}}_2+\underbrace{\textbf{apec}}_3+\underbrace{\textbf{apec}}_4)+\underbrace{\textbf{powerlaw}}_5\quad.
\end{equation}

\begin{figure}
	\centering
	\includegraphics[scale=0.3]{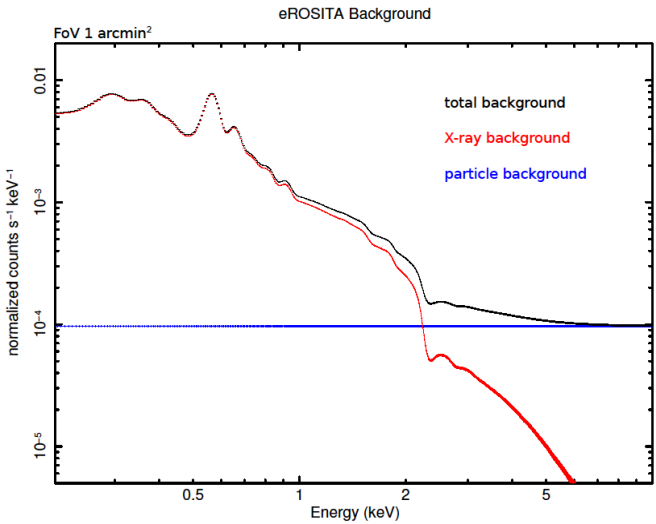}
	\caption{Spectrum of the \textit{eROSITA} background for a FoV of $1$ arcmin$^2$.}
	\label{PixBackground}
\end{figure}

The different components include 1) the absorption by the neutral gas in our Galaxy, 2) the unresolved cosmic X-ray background, i.e. distant AGN, 3) the plasma emission by the hot ISM and 4) the emission by supernova remnants in our Galaxy, as well as 5) the particle background. The first four components are defined by the work by \cite{Lumb2002} and express the cosmic X-ray background, whereas the particle background is estimated by \cite{Tenzer2010}. The instrumental background is included in the particle background and since the \textit{eROSITA} detectors will be equipped with a graded-Z shield, we do not expect to observe a significant component of fluorescent emission lines. Additionally, the influence of bad and hot pixels is assumed to be negligible. The individual values for the model are presented in Table \ref{TabeROSITAbackground}, where all components except the particle background are convolved with the instrumental RSP. This background model is the default for the \textit{eROSITA} instrument and is also described by \cite{Merloni2012}.\\\indent
Figure \ref{PixBackground} provides an illustration of the background spectrum, which is dominated by the particle background for energies above $\sim2$ keV. When observed over the entire \textit{eROSITA} field-of-view (FoV) of $0.83$ deg$^2$, the total background emission shows count rates of $12$ cts/s within the energy range between ($0.3-8$) keV. For a commonly observed cluster of $M_{500}=10^{14}\text{ M}_{\odot}$ ad $z=0.1$ as simulated in Fig. \ref{PixClusterSpectrum}, this background results in a signal-to-noise ratio of $S/N\approx23.5$ and in a source-to-background ratio of $1.4$.

\begin{table} [tb]
	\centering
	\begin{tabular}{c|c|c}\hline\hline
		\multirow{2}{*}{Component}	& \multirow{2}{*}{Parameter}	& \multirow{2}{*}{Model Value} 							\\
									&								&		\\\hline
		1					& $N_{\text{H}}$ & $1.7\times10^{-2}$		\\
		2					& photon spectral index				& 1.42 							\\
		2					& norm								& 0.0028	 \\
		3					&	k$T$								&	0.204 		\\	
		3					&	norm							&	0.0019	 \\
		4					&	k$T$									&	$7.4\times10^{-2}$\\
		4					&	norm							&	0.029 \\
		5					&	photon spectral index 	& 0.0					\\
		5					&	norm							&	0.29 \\\hline
	\end{tabular}
	\caption{Model values of the \textit{eROSITA} background. The numbering of the components is equivalent to the numbering in the model definition (Sec. \ref{background}). The units of the individual model parameters are as follows: [$N_{\text{H}}$]= $10^{22}$ particles/cm$^2$ and [k$T$]=keV. The normalisations are given for an \textit{eROSITA} field-of-view of $0.83\text{ deg}^2$ with the units [norm]=photons/keV/cm$^2$/s at $1$ keV for the powerlaw and [norm]=photons/cm$^5$ for the apec model.}
	\label{TabeROSITAbackground}
\end{table} 

\subsection{Simulation outline}
\label{SimOutline}

To simulate the characteristics of \textit{eROSITA} galaxy clusters the following methodology is applied:
\begin{enumerate}
	\item{For a given set of cluster mass and redshift, we simulate the total X-ray spectrum, which includes both the absorbed galaxy cluster emission itself as well as the background.}
	\item{A model is fit to the simulated emission. However, before the fitting procedure, we define the background emission, such that this emission is removed from the above spectrum during the fit and only the model of an absorbed cluster emission needs to be adjusted to the remaining spectrum. The fit then determines the best fit values of the cluster temperature and redshift.}
	\item{To obtain a proper statistical distribution of these best-fit values, steps 1.) - 3.) are repeated 300 times for each parameter set.}
\end{enumerate}
For the simulations we define two different exposure times $t_{\text{exp}}=1.6$ ks and $t_{\text{exp}}=20$ ks, which describe the effective average exposure time for \textit{eROSITA} after its four years of all-sky surveys and the observation time of two deep exposure fields at the ecliptic poles, respectively \citep{Pillepich2011,Merloni2012}. Also for the fitting we follow two approaches, which assume the redshift either to be known, e.g. from optical follow-up observations, or that no redshift information is available yet. In the latter case, we introduce the redshift as variable parameter during the fit and determine its value through the X-ray spectrum \citep[e.g.,][]{Yu2011}. These considerations yield a total of four different simulations.\\\indent
Throughout the different simulation steps, Cash statistics are applied \citep{Cash1979} to ensure a good performance during the fit despite the small number of photons in each energy bin of the simulated spectra. For the realisation of the total spectrum during the first step, we define the exposure time and convolve the emission models of the cluster and of the background with the instrumental responses, where the background normalisation is rescaled to match the cluster extension (see Sec. \ref{ClusterProp}). The spectrum is re-grouped to yield at least one photon count per energy bin, to avoid failures during the fit due to empty energy bins \citep[compare][]{Leccardi2007,Krumpe2008}. In the next step, the background emission is defined by applying the {\tt{backgrnd}}-command, for this emission to be removed during the final spectral fit. The procedures of normalising the background and employing the {\tt{backgrnd}}-command are essential to account for the statistical scatter in the photon counts in the spectra. For this background model, we realistically assume an exposure time of $t_{\text{exp}}=100$, while keeping the area fixed to the cluster extension. Finally, an absorbed apec emission model is fit to the remaining spectrum within an energy range of ($0.3-8$) keV, which reflects the effective energy range of the \textit{eROSITA} instrument \citep{Merloni2012}. During this simulation step, the cluster temperature and the normalisation of the spectrum, which is proportional to the emission measure 
\begin{equation}
	\text{norm}=\frac{10^{14}}{4\pi[D_{\text{A}}(1+z)]^2}\int n_{\text{e}}N_{\text{H}}\text{d}V\quad ,
\end{equation}
are recovered. In case of an unavailable cluster redshift, also this property is estimated in this step.\\\indent
To allow for the most accurate fit values to be obtained, we thoroughly inspect the more-dimensional space of the best-fit parameters for a global minimum in the goodness of the fit by applying the multi-dimensional {\tt{steppar}}-command. The investigated parameter space is defined as $\pm50\%$ around the initial best-fit value with $50$ steps each for the temperature and the redshift and $20$ steps for the normalisation. In a last step, we estimate the $68\%$-confidence intervals of the best fit values by means of the \textit{xspec} {\tt{error}}-command. The complete simulation procedure is then repeated $300$ times for each set of parameters resulting in a well-sampled distribution of best-fit values. This distribution allows us to define a second $68\%$-confidence interval around the median best-fit value. In the following, this last confidence range is applied for the analysis of the simulation and is considered as true uncertainty on the fit values.

\subsection{Analysis procedure}
\label{Analysis}

Before the analysis of the simulated data, we remove all catastrophic failures in the fit results, which we devide into two types. The first type of catastrophic failures contains inconsistencies in the fit, where the $68\%$-confidence interval calculated by the {\tt{error}}-command is not set around the best fit value. These inconsistencies may occur during the simulations with unknown redshift. When appearing in the analysis of observed data, the spectral fit needs to be repeated, while being adapted individually to this spectrum by means of e.g. re-defined starting values for the fit. This approach is not feasable for the extend of our simulations, such that we are limited to the conservative procedure of discarding these spectra.\\\indent
During the analysis, we address each parameter set separately and define the second type of catastrophic failures as fit values, whose true $3\times 68\%$-confidence interval does not include the input value. This type of failures can only be quantified if the input cluster parameter values are known. In the analysis of observed data, however, they cannot be identified and thus decrease the accuracy of the analysed data sample. \\
If both types of catastrophic failures sum up to more than $20\%$ of the fit data, the parameter set is rejected (see Sec. \ref{Discussion}), i.e. it is assumed that the cluster property values cannot be recovered typically from the \textit{eROSITA} data. However, to ensure a conservative analysis, those fits showing the second type of catastrophic failure are included in the analysis of all our data sets since these catastrophic failures can generally not be identified for observed data.\\\indent
The analysis considers three different interpretations of the temperature and the redshift fit results, all of which are presented in dependence on the input values of the cluster mass and redshift. First we inspect the relative uncertainties, which we define as $\Delta T/\langle T_{\text{fit}}\rangle$ and $\Delta z/\langle 1+z_{\text{fit}}\rangle$, respectively. The elements $\Delta T$ and $\Delta z$ express the true $68\%$-confidence range from the distribution of the fit values. The typical fit values $\langle T_{\text{fit}}\rangle$ and $\langle1+z_{\text{fit}}\rangle$ are estimated by the median of the distribution. Especially of interest are relative uncertainties of both properties with values of $\lesssim10\%$ since these uncertainties are comparable to the intrinsic scatter in the  $M$-$T$ scaling relation \citep[e.g.,][]{Mantz2010}. We emphasise on the fit results of the temperature since for future \textit{eROSITA} observations the total cluster mass is more precisely estimated by the $M-T$ relation, due to its smaller intrinsic scatter compared to the $M-L$ relation \citep[e.g.,][]{Mittal2011}. However, the analysis of the recovery of the cluster mass from the simulated spectra is beyond the scope of this paper.\\\indent
The bias on the best fit cluster properties is computed as $\langle T_{\text{fit}}\rangle/T_{\text{input}}$ and $\langle1+z_{\text{fit}}\rangle/(1+z_{\text{input}})$, respectively, expressing the ratio between the median of the fit values and the input value. As a last analysis, we investigate the deviation between the median uncertainty computed by the {\tt{error}}-command and the uncertainty obtained from the distribution of the fit results as $\langle\Delta T_{\text{error}}\rangle/\Delta T$. Analogously, the deviation in the redshift uncertainties is analysed. This so-called bias in the error estimates is an important quantity since from the reduction of observed data only the uncertainty by the {\tt{error}}-command will be available, whereas the proper statistical uncertainty is given by the distribution.


\section{Results}
\label{Results}

\subsection{Relative uncertainties}
\label{RelUncertain}

\begin{figure}[H]
	\centering
	\includegraphics[scale=0.37,trim=0cm 5cm 0cm 4cm]{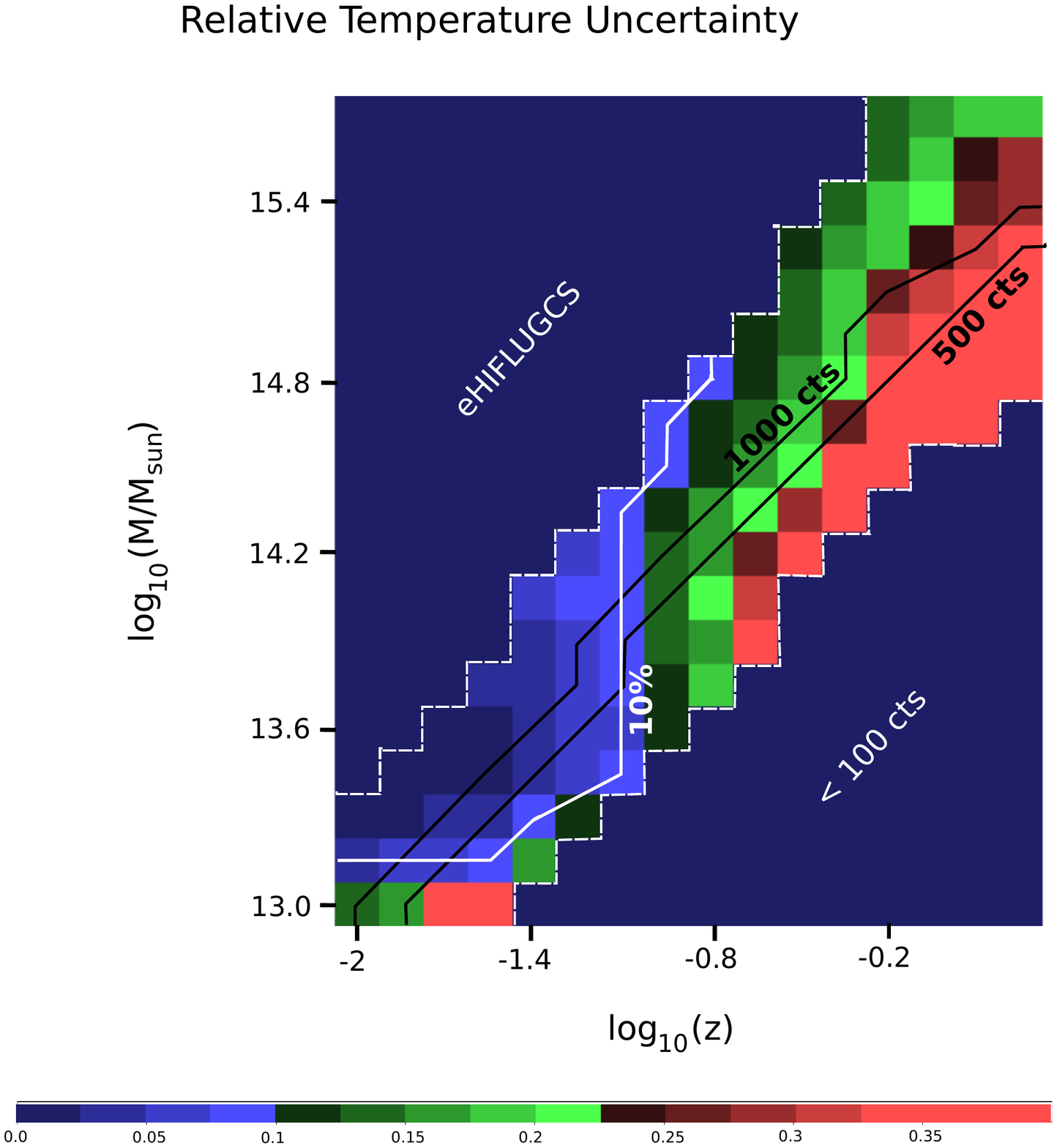}
		\caption{Expected relative temperature uncertainties $\Delta T/\langle T_{\text{fit}}\rangle$ in dependence on the total cluster mass and the cluster redshift. This simulation assumes an exposure time of $t_{\text{exp}}=1.6$ ks and the redshift of the clusters to be available. We present white and black contour lines to emphasise on the relative uncertainties and the number of source counts, respectively.}
		\label{Pix16ksRelT}
	\includegraphics[scale=0.37,trim=0cm 8cm 0cm 5cm]{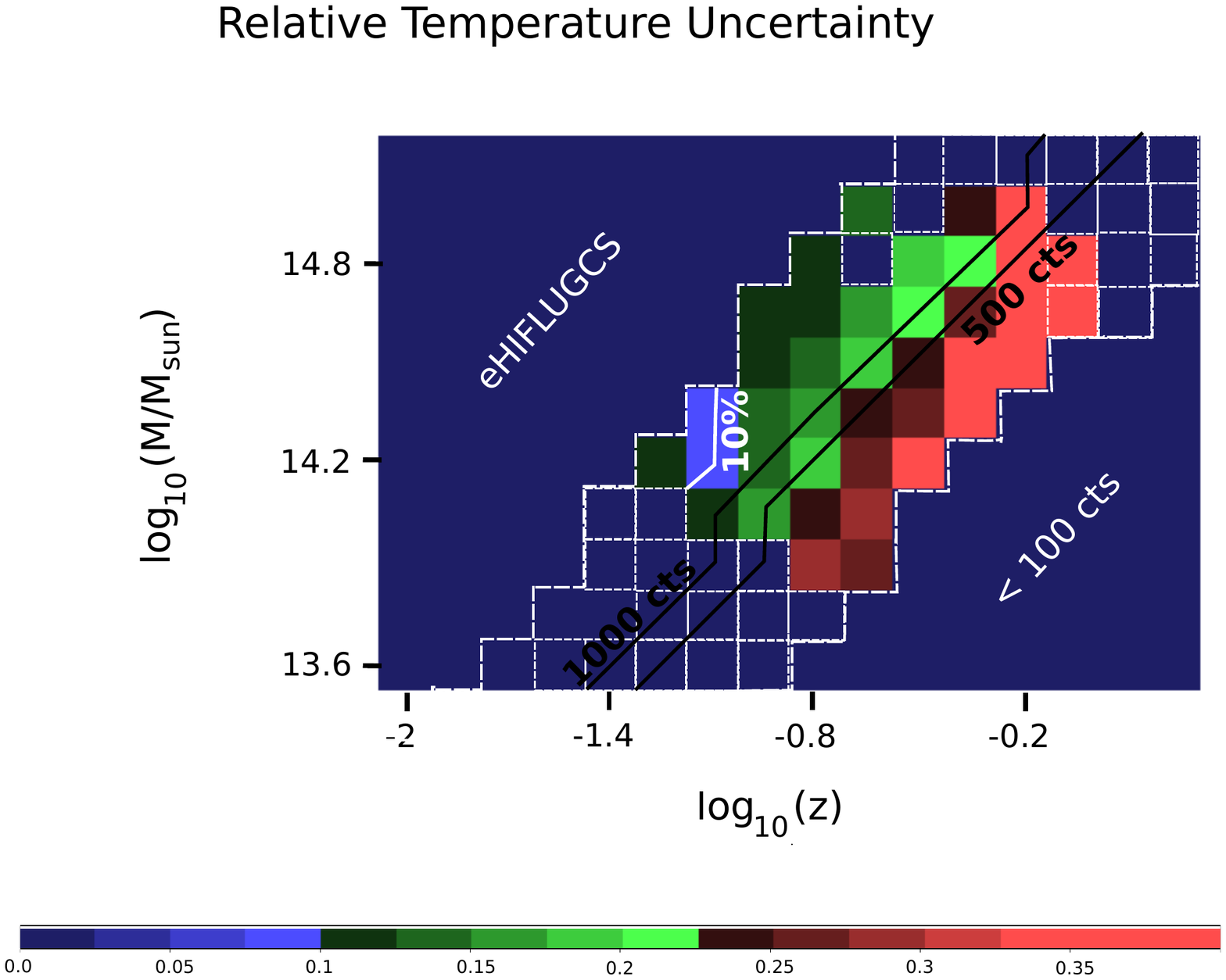}
		\caption{Relative temperature uncertainty as presented in Fig. \ref{Pix16ksRelT}, but assuming the cluster redshift to be unavailable. The dark blue pixels with dashed white contours indicate the catastrophic failures in the fit.}
		\label{Pix16ksVarZRelT}
	\includegraphics[scale=0.37,trim=0cm 8cm 0cm 4.2cm]{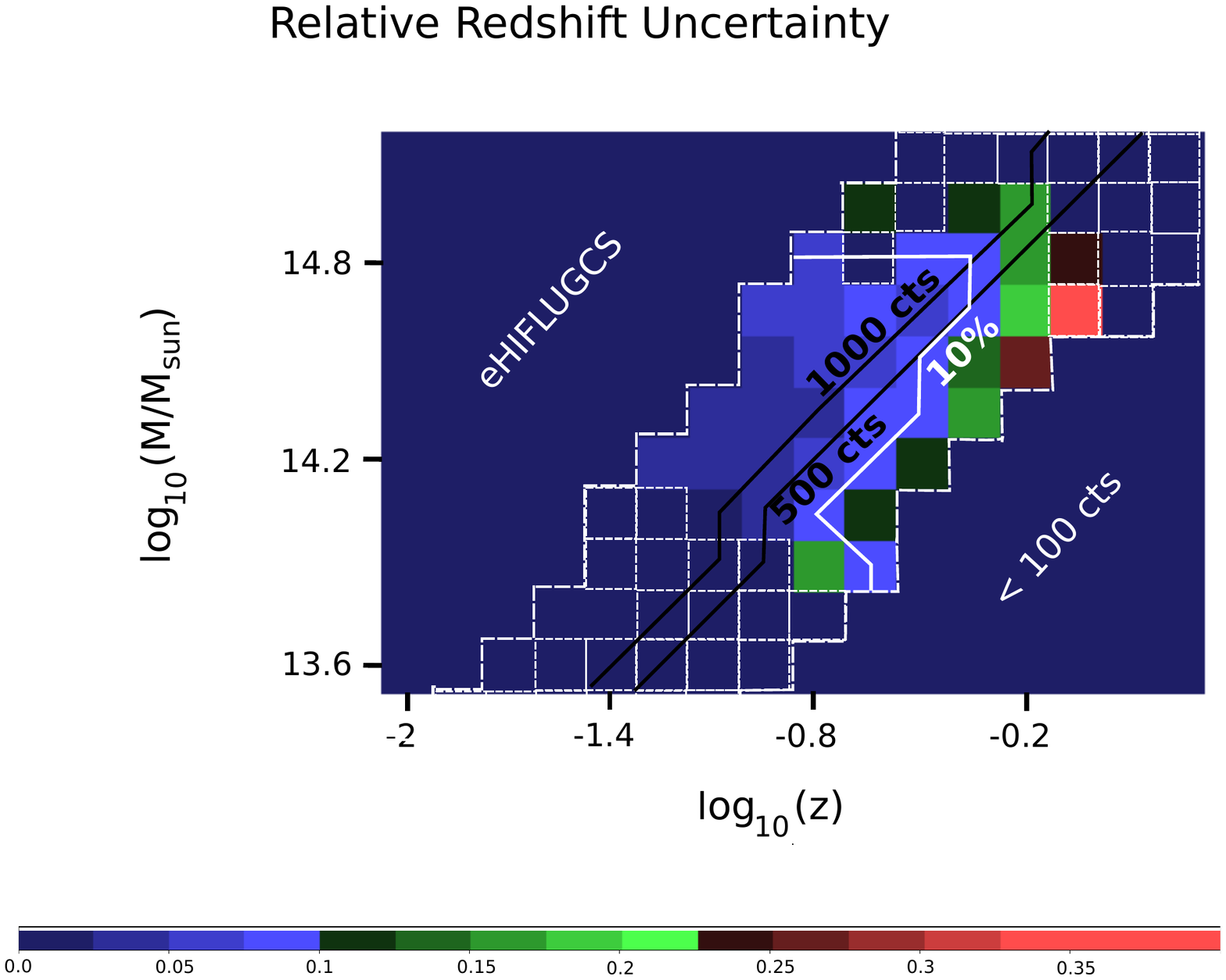}
		\caption{Expected relative redshift uncertainty $\Delta z/\langle1+z_{\text{fit}}\rangle$ for an exposure time of $t_{\text{exp}}=1.6$ ks.}
		\label{Pix16ksVarZRelZ}
\end{figure}

\begin{figure}[H]
	\centering
	\includegraphics[scale=0.37,trim=0cm 5cm 0cm 4cm]{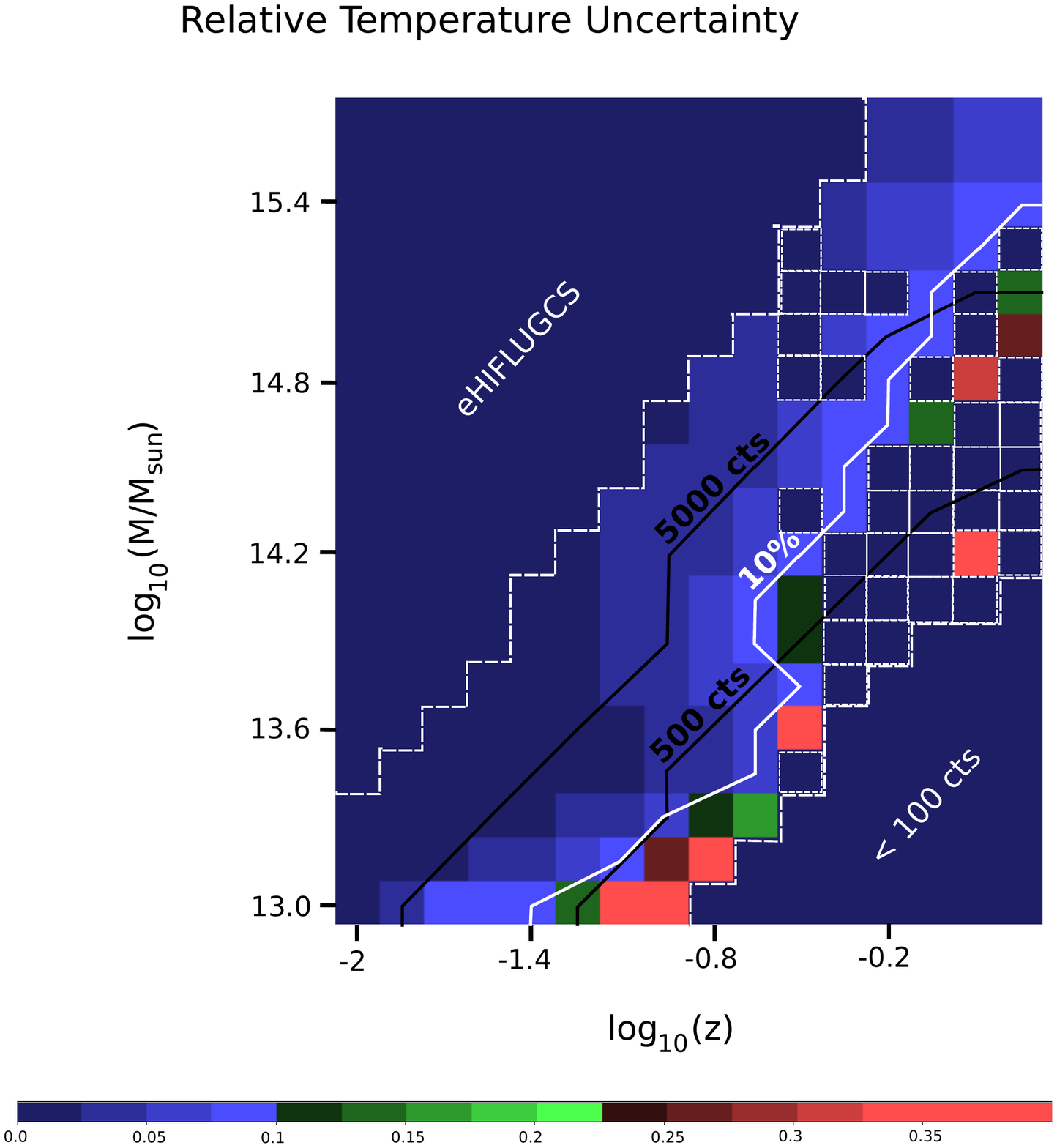}
		\caption{Expected relative temperature uncertainty for an exposure time of $t_{\text{exp}}=20$ ks, if the redshift of the cluster is available.}\vspace*{0.5cm}
		\label{Pix20ksRelT}\vspace*{-0.1cm}
	\includegraphics[scale=0.37,trim=0cm 8cm 0cm 2.3cm]{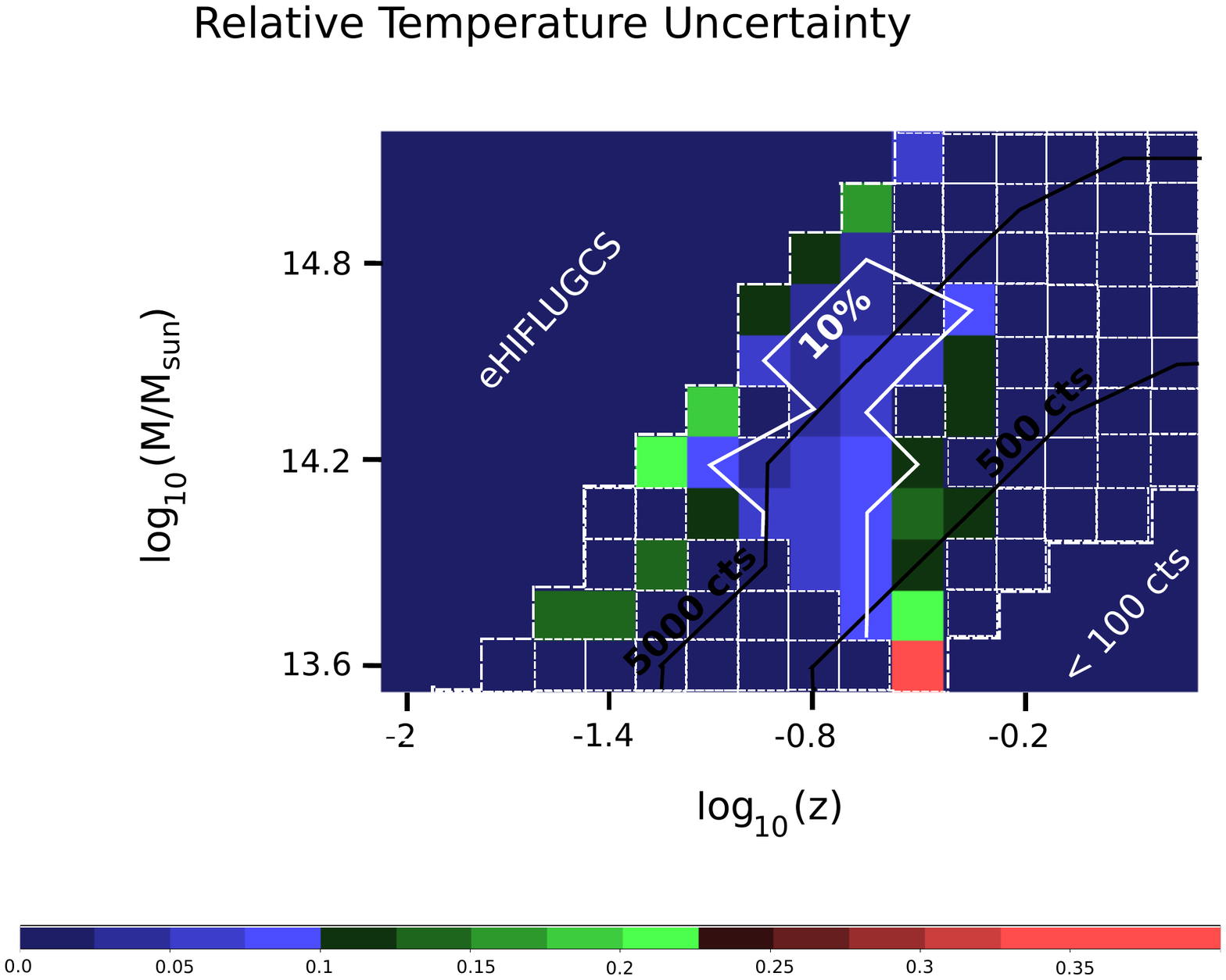}
		\caption{Expected relative temperature uncertainty for the simulation of $t_{\text{exp}}=20$ ks and unknown cluster redshift.}\vspace*{0.3cm}
		\label{Pix20ksVarZRelT}
	\includegraphics[scale=0.37,trim=0cm 8cm 0cm 4cm]{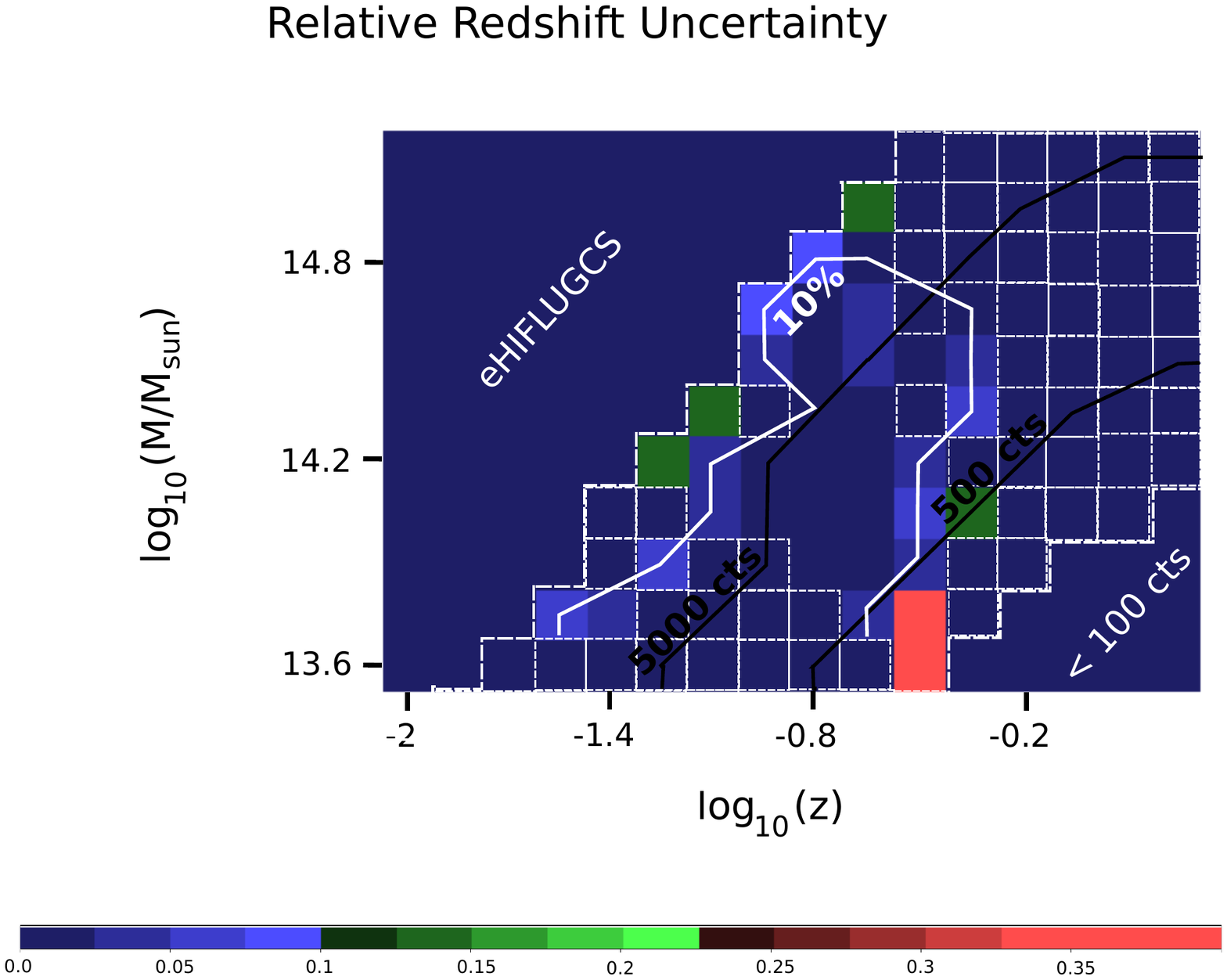}
		\caption{Expected relative redshift uncertainty for an exposure time of $t_{\text{exp}}=20$ ks.}
		\label{Pix20ksVarZRelZ}
\end{figure}

Figures \ref{Pix16ksRelT} to \ref{Pix20ksVarZRelZ} illustrate the relative temperature and relative redshift uncertainties, which are expected after the four years of \textit{eROSITA} all-sky survey. The relative uncertainties are computed in dependence on the input cluster mass and redshift, such that each pixel represents a galaxy cluster with a different combination of input mass and redshift, where the values of the two properties are given by the centre of the pixel. The colour of the pixel indicates the relative uncertainty of either the temperature or the redshift of the cluster. The colour bar expresses this relative uncertainty and is given in a linear scale. According to the defined flux limit and photon count limit (Sec. \ref{SimOutline}), only the cluster parameter space within the two white dashed lines is considered. In the simulation of the \textit{eROSITA} deep exposure fields with $t_{\text{exp}}=20$ ks, this parameter space increases to higher redshifts as fainter clusters are detected above the photon count threshhold (Fig. \ref{Pix20ksRelT} to \ref{Pix20ksVarZRelZ}).\\\indent
For display purposes, we include countour lines for the relative uncertainty in white and for the number of detected photons in black, where each cluster on the contour line shows at least the stated precision or number of photons. Within figures \ref{Pix20ksVarZRelT} and \ref{Pix20ksVarZRelZ}, the parameter space of clusters with relative uncertainties of $\lesssim10\%$ in temperature or redshift is indicated as area between the solid white contour lines. The white-framed dark blue pixels present the rejected parameter sets due to a large fraction of catastrophic failures (Sec. \ref{Analysis}). \\
\\\indent
In comparison to the simulation with $t_{\text{exp}}=1.6$ ks and known redshift (Fig. \ref{Pix16ksRelT}), the number of rejected pixels increases if the exposure time is increased and especially if we assume the redshift to be not available. For the simulation results with unknown redshift the figures are clipped to the intermediate mass range of $13.6\lesssim\log(M/\text{M}_{\odot})\lesssim15.1$ since all parameter sets including the remaining masses are rejected (Figs. \ref{Pix16ksVarZRelT} $\&$ \ref{Pix16ksVarZRelZ} and \ref{Pix20ksVarZRelT} $\&$ \ref{Pix20ksVarZRelZ}). With increasing exposure time, the increased number of detected photons reduces the statistical scatter in the simulated spectra, which allows for a higher precision of the fit. Accordingly, this raised precision tightens the absolute constraints on the catastrophic failures. Futhermore, the introduction of the redshift as additional free parameter in the simulations complicates the fitting procedure and yields less accurate and less precise fit results (Sec. \ref{ParamBias}). The occurence of a high level of failed spectral fits when determining the X-ray redshift of a cluster has as well been observed by \cite{Lloyd2011} (see also Sec. \ref{Failures}).\\
In all simulation approaches, the precision of both temperature and redshift generally increases with increasing cluster mass and, in particular, with decreasing cluster redshift. \\\indent
According to these findings, the galaxy clusters, which are relevant for cosmological studies with relative parameter uncertainties of $\leqslant10\%$, are observed in the local universe. For the all-sky survey with an average effective exposure time of $t_{\text{exp}}=1.6$ ks, we expect the temperature to be detectable with this precision up to maximum redshifts of $\log(z)\approx-0.8$, $z\lesssim0.16$ (Fig. \ref{Pix16ksRelT}), if the redshift of the cluster is known, and up to $\log(z)\approx -1.1$, $z\lesssim0.08$ (Fig. \ref{Pix16ksVarZRelT}), if the redshift is not available. The redshift itself will be ontained with relative uncertainties of $\leqslant10\%$ from X-ray data for clusters as far as $\log(z)\approx-0.35$, $z\lesssim0.45$ (Fig. \ref{Pix16ksVarZRelZ}). At the ecliptic poles of the mission with exposure times of $t_{\text{exp}}=20$ ks, the parameter space of clusters with precision temperatures increases in theory up to redshifts of $z\lesssim1.78$ (Fig. \ref{Pix20ksRelT}), assuming the redshift to be known. At these redshifts precise temperatures are only obtained for the most massive galaxy clusters of which not many are expected to be observed (compare Fig. \ref{PixDistribution}), especially in the low sky area of the deep exposures. Additionally, pollution of the spectra by the cluster AGN needs to be expected for these deep observations (see Sec. \ref{RemarksNumbers}). In the case of unavailable redshifts, both temperature and redshift are detectable up to $\log(z)\approx-0.35$, $z\lesssim0.45$ (Figs. \ref{Pix20ksVarZRelT} $\&$ \ref{Pix20ksVarZRelZ}). For these observations, catastrophic failures in the spectral fit restrict the parameter space of clusters with precise temperature and redshift estimates.\\\indent
The parameter space of clusters with high precision temperatures decreases for the simulation with unknown redshift, due to the introduction of the redshift as additonal free parameter during the fit and the resulting degeneracy between the cluster redshift and the cluster temperature (compare Sec. \ref{Failures}). In comparison to the cluster temperature, the cluster redshift is more difficult to determine from X-ray spectra \citep[e.g.,][]{Yu2011,Lloyd2011}. Only because of the deviating definitions of the relative uncertainties as $\Delta T/\langle T_{\text{fit}}\rangle$ and $\Delta z/\langle 1+z_{\text{fit}}\rangle$, precise redshifts are expected to be detected for more distant clusters in comparison to precise temperatures. According to this, the number of clusters for which both precise redshifts and temperatures will be available from X-ray data is limited by the determination of the temperature.\\\indent 
The analysis of the relative uncertainties shows clearly that the precision of both temperature and redshift does not only depend upon the number of detected photons, but also upon the cluster properties themselves (see Sec. \ref{DependenceUncertain}).

\begin{figure}[H]
	\centering
	\includegraphics[scale=0.35,trim=0.5cm 2cm 1cm 2cm]{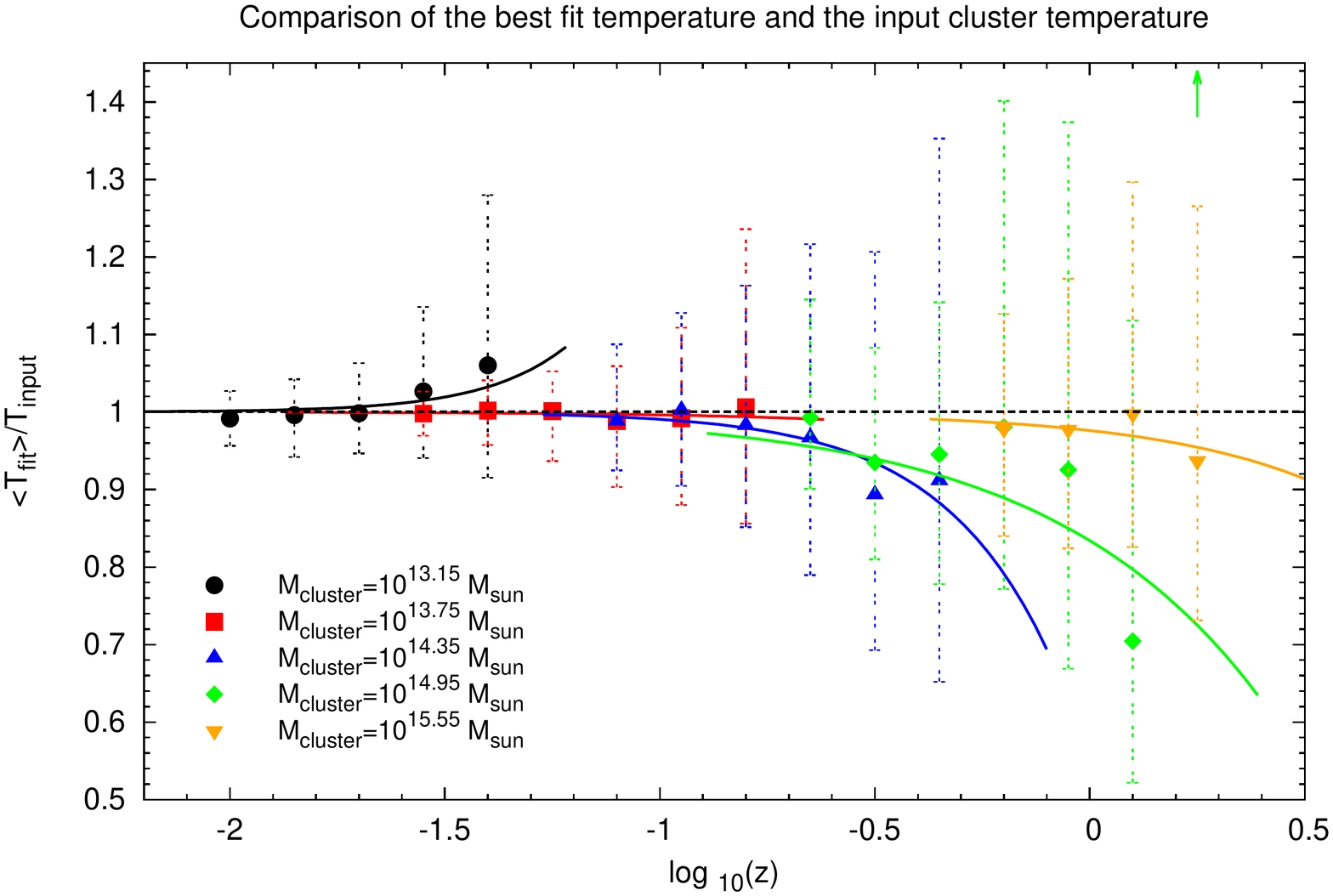}
		\caption{Bias on the best-fit temperature in dependence on the cluster redshift for the simulation of $t_{\text{exp}}=1.6$ ks and known redshift. For each displayed cluster mass individual bias correction functions are suggested as solid curves with the corresponding colour. }
		\label{PixBias16ks}
	\includegraphics[scale=0.35,trim=0.5cm 2cm 1cm 1cm]{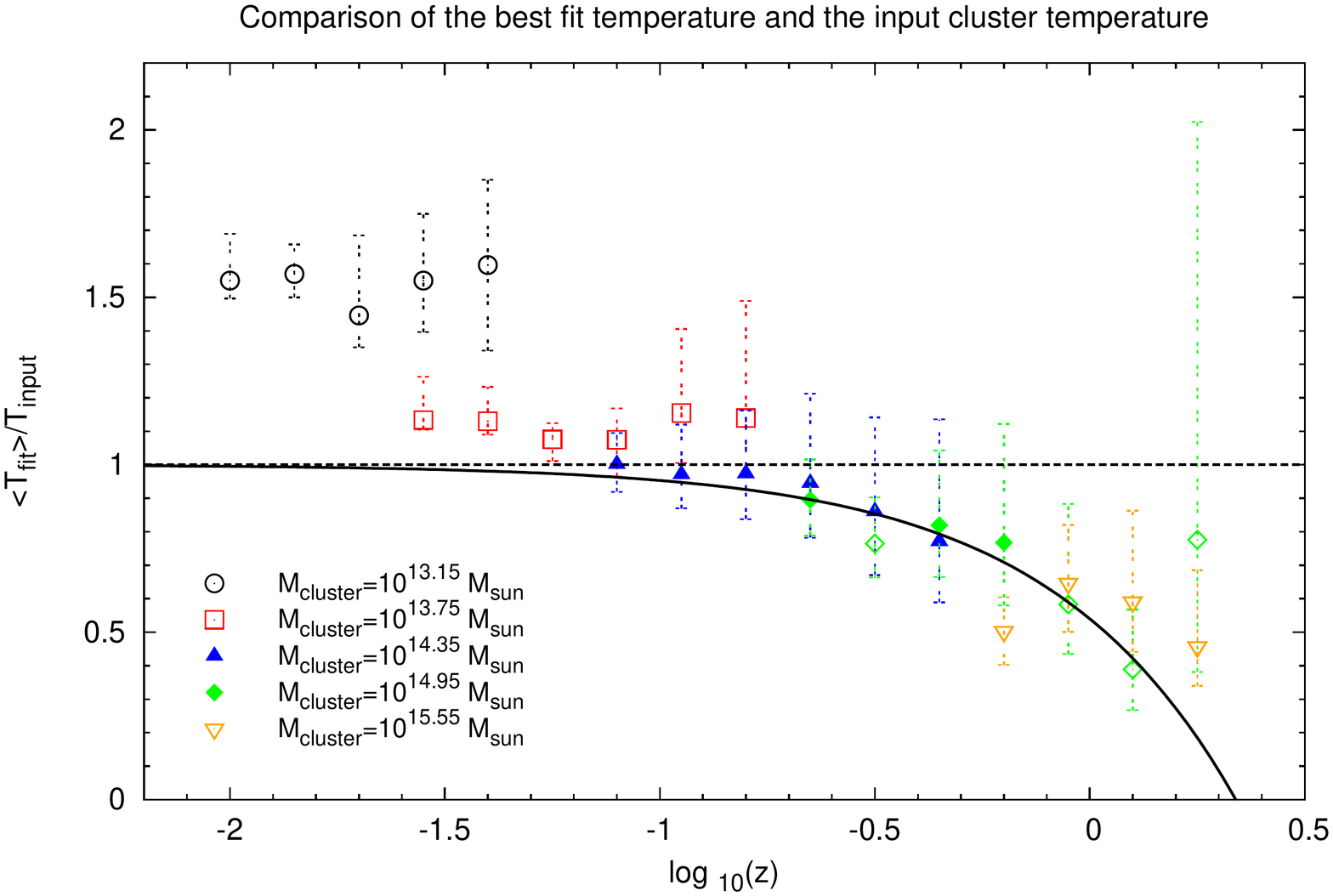}
		\caption{Bias on the best-fit temperature for the simulation of $t_{\text{exp}}=1.6$ and assuming the cluster redshift to be unavailable. For all simulated cluster masses the bias is described by a single function, which is presented as solid black curve, where the empty symbols indicate the rejected data sets due to a large fraction of catastrophic failures.}
		\label{PixBiasT16ks}
	\includegraphics[scale=0.35,trim=0.5cm 2cm 1cm 1cm]{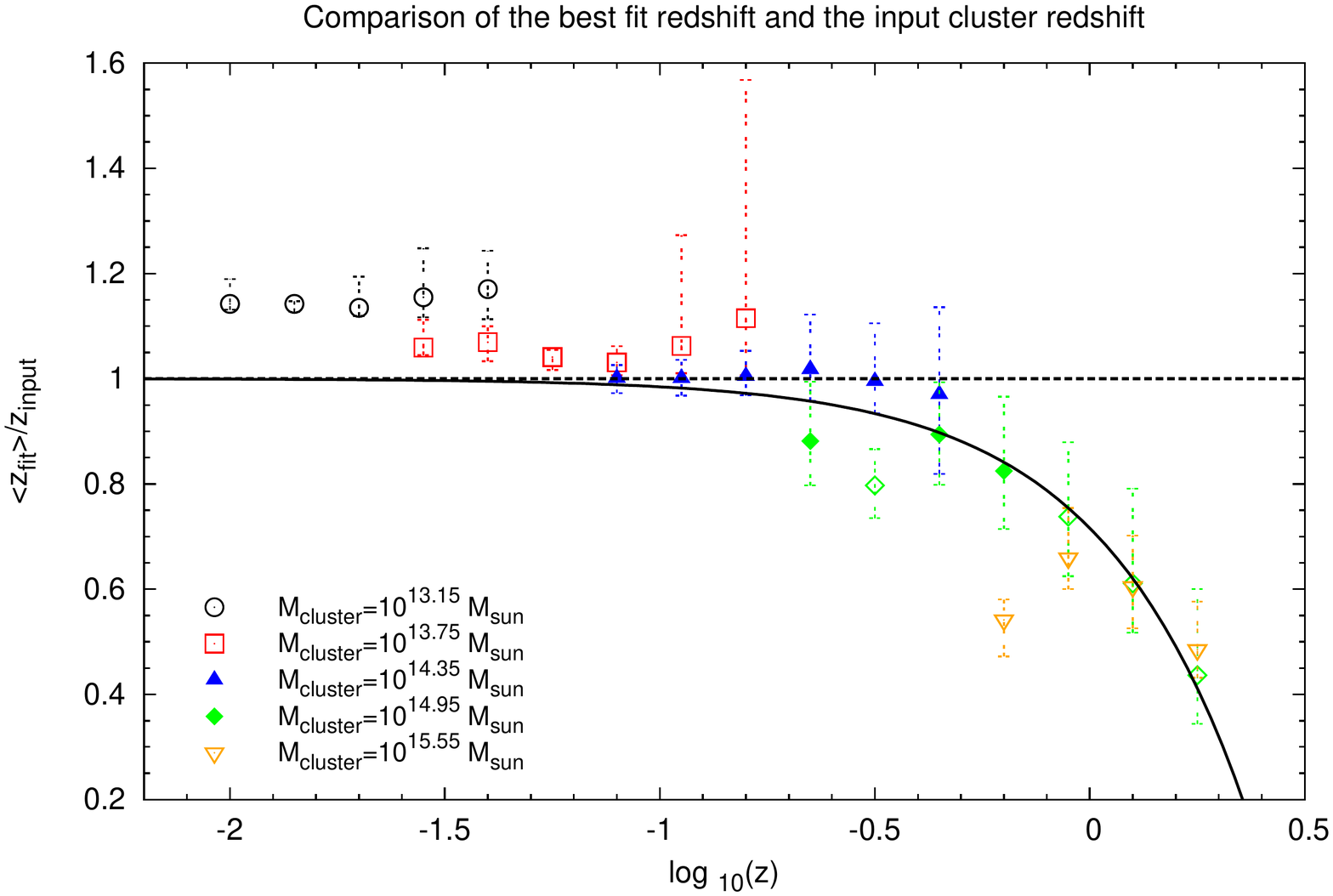}
		\caption{Bias on the best-fit redshift assuming an exposure time of $t_{\text{exp}}=1.6$. Also the bias on the redshift can be parameterised by a single function for all cluster masses.}
		\label{PixBiasZ16ks}
\end{figure}
\begin{figure}[H]
	\includegraphics[scale=0.35,trim=0.5cm 2cm 0.5cm 2cm]{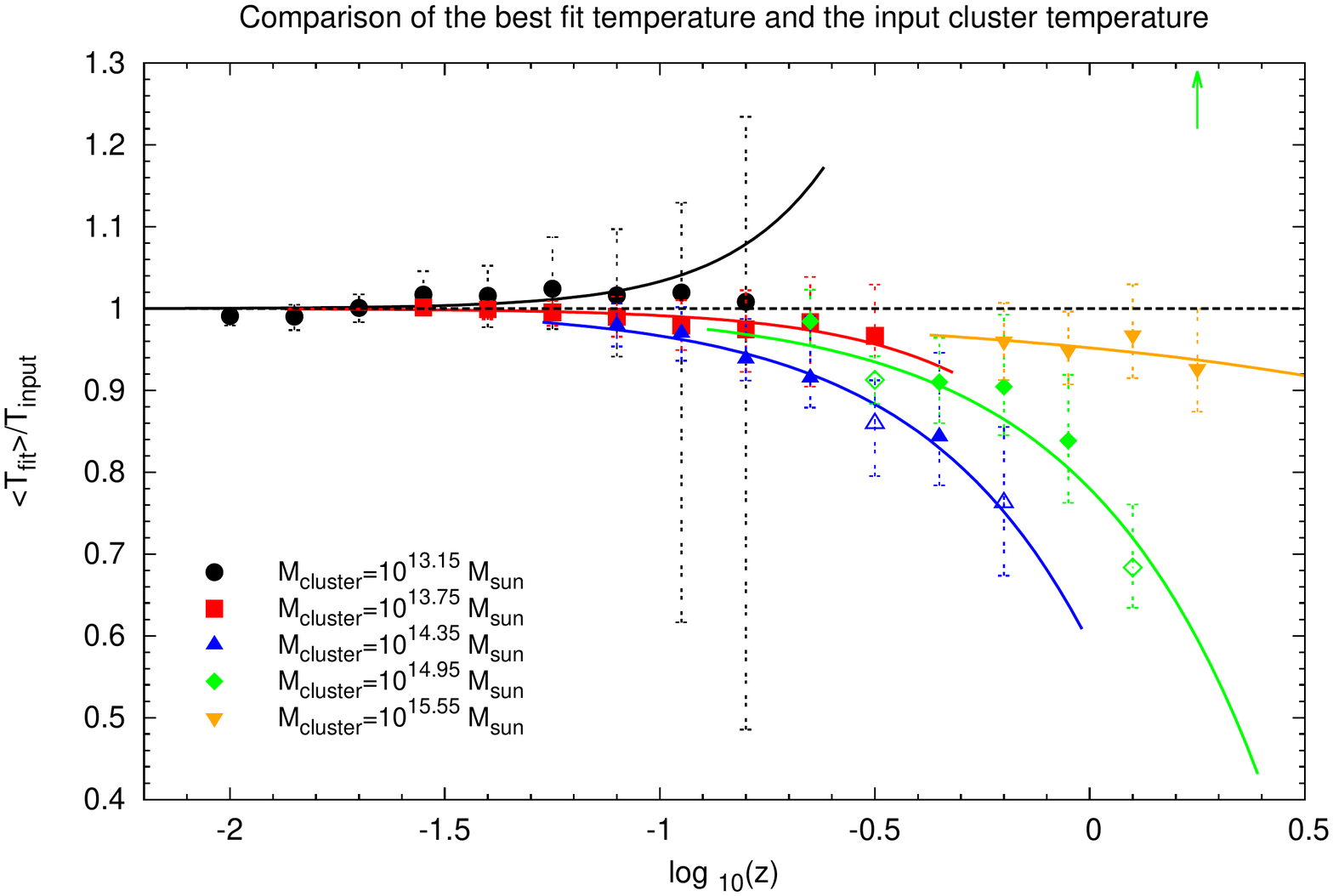}
		\caption{Bias on the best-fit temperature for the deep exposure fields of $t_{\text{exp}}=20$ ks and for clusters with known redshift. Again, suggested correction functions for this bias are presented.}
		\label{PixBias20ks}
	\includegraphics[scale=0.35,trim=0.5cm 2cm 0.5cm 0cm]{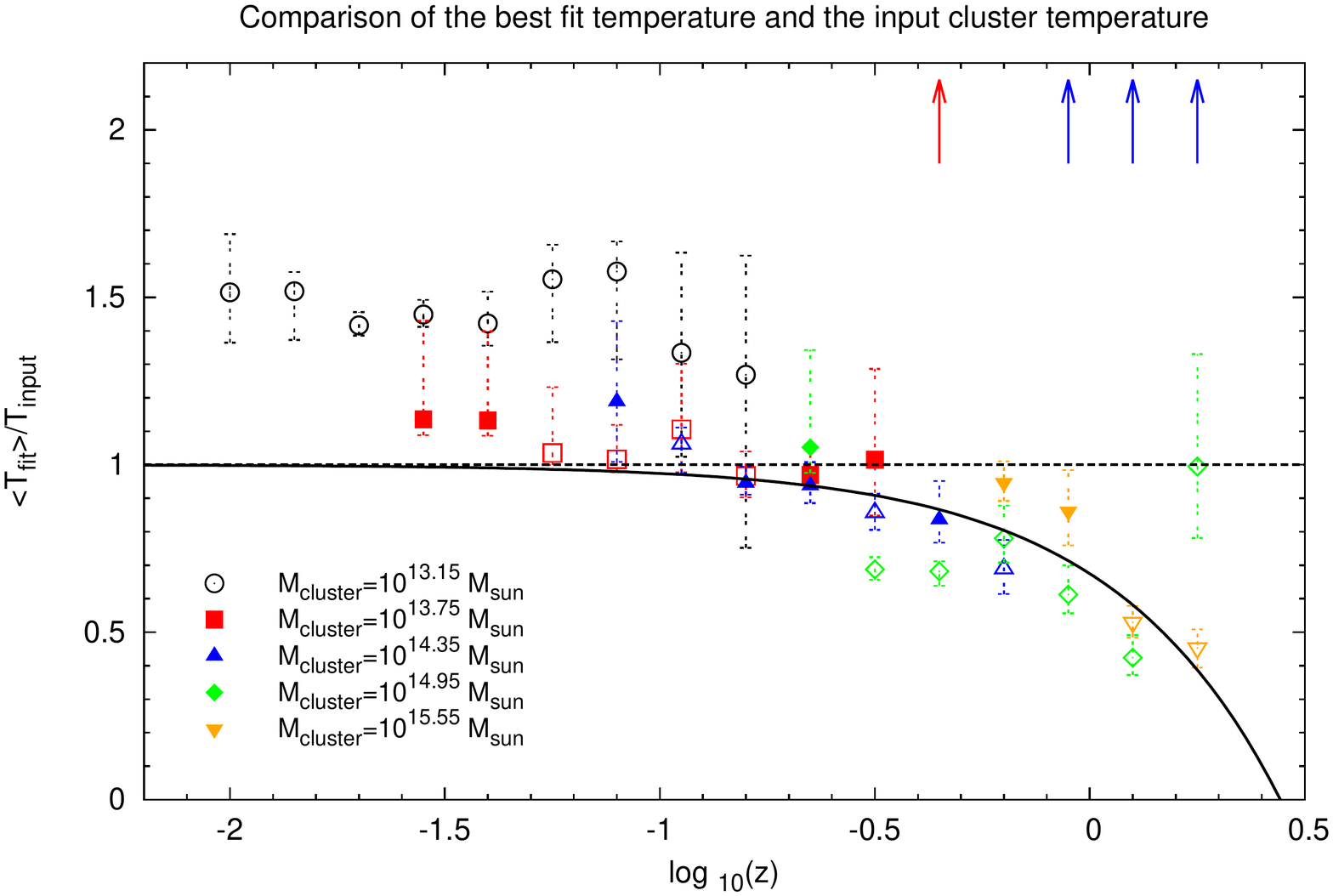}
		\caption{Bias on the best-fit temperature for clusters in the deep exposure fields with unknown redshift. For the entire mass range, the bias is described by a single function.}
		\label{PixBiasT20ks}
	\includegraphics[scale=0.35,trim=0.5cm 2cm 0.5cm -1.0cm]{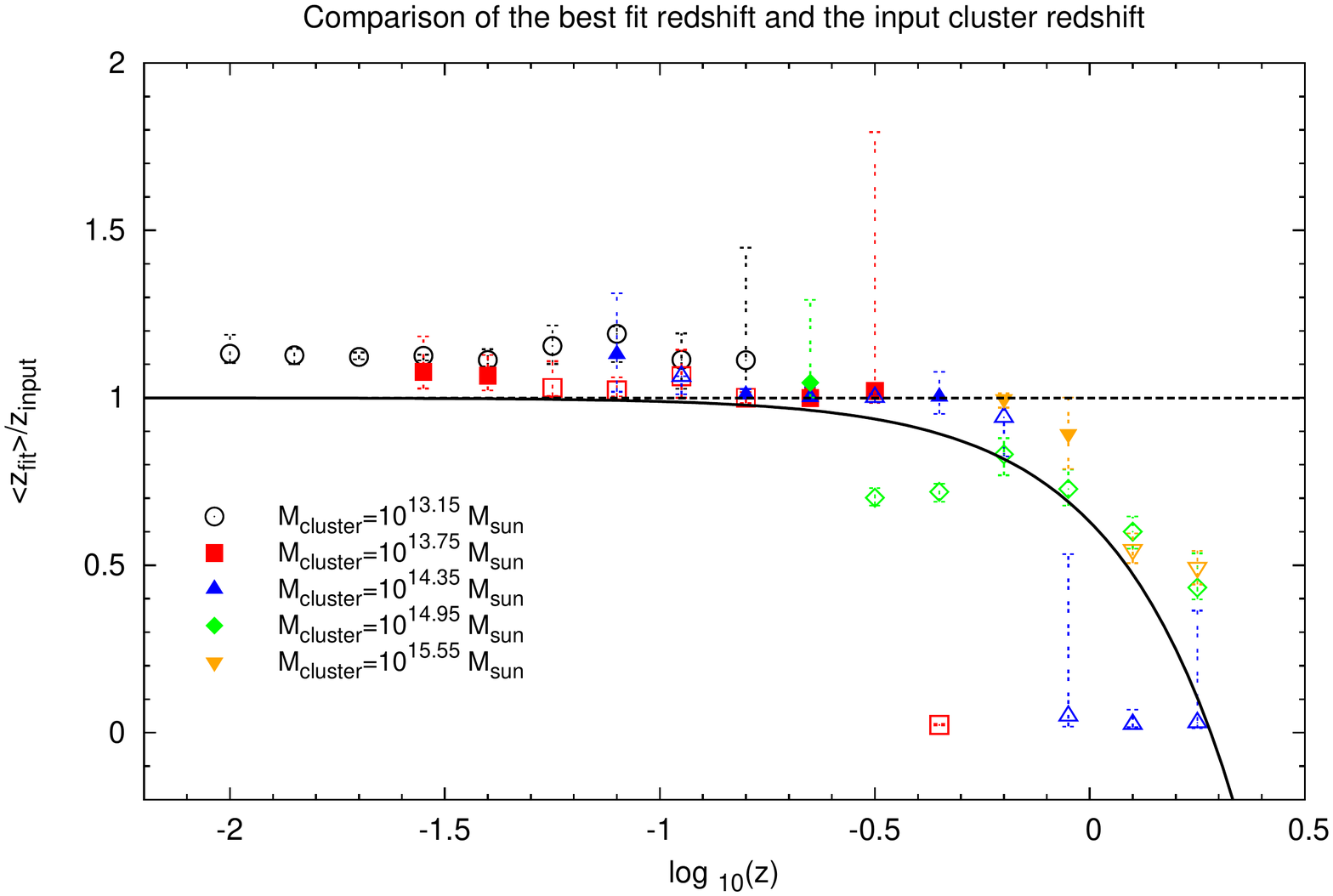}
		\caption{Bias on the best-fit redshift for clusters in the deep exposure fields and the estimated correction function for this bias.}
		\label{PixBiasZ20ks}
\end{figure}
\newpage

\subsection{Biases in the best-fit properties}
\label{ParamBias}
The bias in the best fit temperatures and redshifts is analysed in dependence on the cluster redshift within five mass ranges, defined by the input cluster masses. These mass intervals are centered around the values $\log(M/\text{M}_{\odot})=13.15$, $13.75$, $14.35$, $15.95$ and $15.55$, where the parameter biases of these cluster masses are illustrated in figures \ref{PixBias16ks} - \ref{PixBiasZ20ks} within the simulated redshift intervals. The uncertainty of the bias is given by the scatter in the best-fit values. We also present correction functions for these biases, which we obtain as fit of the exponential function
\begin{equation} 
f(x)=A\cdot\exp(B\cdot x)+1\quad ,
\label{EqExpFit}
\end{equation} 
with variables $A$ and $B$ and $x=\log(z)$, to the data points.\\\indent
The best fit values of $A$ and $B$ are provided in the appendix (Appendix \ref{AParameterBias}). The parameter sets, which are rejected due to large numbers of catastrophic failures, are displayed as empty symbols. They are included in the fit of the correction function to avoid an underestimation of the correction of the best-fit property values. However, those cluster masses, which show only catastrophic failures for all redshifts, are excluded from this fit. For the simulations with known redshift, we define correction functions individually for the five cluster masses stated above. However, we assume the correction function to be an estimate for all masses within the defined mass range and within the simulated redshift interval (Appendix \ref{AParameterBias}). If the cluster redshift is unknown, the parameter biases are to a first approximation independent of the cluster mass (Figs. \ref{PixBiasT16ks} $\&$ \ref{PixBiasZ16ks} and \ref{PixBiasT20ks} $\&$ \ref{PixBiasZ20ks}). According to this, we describe these biases by a single exponential function for all cluster masses. The degeneracy in the cluster masses occurs as for the simulation with unavailable redshift a greater scatter is introduced in the median values of the parameter bias.\\\indent
In general, the biases in the best-fit properties exhibit a decrease with declining cluster redshift and for the simulated clusters with known redshifts, the bias additionally increases with decreasing cluster mass. For local redshifts of roughly $\log(z)\approx-0.7$, the parameter bias becomes negligible for all cluster masses and simulation approaches. Even for higher redshifts the best-fit value is still consistent with the input value within the error bars.\\\indent
With increasing exposure time, the median bias values improve moderately, whereas the uncertainty on the best-fit value decreases significantly. According to this, the bias is only consistent with unity for smaller redshift ranges when compared to the results for $t_{\text{exp}}=1.6$ ks (compare Figs. \ref{PixBias16ks} $\&$ \ref{PixBias20ks}). Similar to the findings for the relative uncertainties, the temperature bias rises if the redshift of the cluster is unavailable. According to the deviating definitions for the temperature and the redshift (Sec. \ref{Analysis}), the redshift appears as more accurate property.\\\indent
The development of the bias in the best-fit properties in dependence on the cluster redshift, the temperature and the number of photons is analogous to the evolution of the relative uncertainties. Thus, both results are explained by similar considerations (see Sec. \ref{DependenceUncertain}). Recall that we investigate an isothermal cluster emission model in our simulations to focus only on the performance of the \textit{eROSITA} instrument. For the analysis of observed data and thus of mainly multi-temperature gas, additional systematics might arise in the temperature estimation, according to the shape of the effective area. A first assessment of this effect is presented by \cite{Reiprich2013} in their Figure 18.\\
The underestimation of the proper, input property value has also been studied by \cite{Leccardi2007}. They explain the deviation through the increasing relative background contribution with increasing redshift in comparison to the source counts as well as through the calibration of the instrument.\\\indent
When convolving these results for the bias in the properties with the parameter space of \textit{eROSITA} clusters with precise temperatures and redshifts, we find that the bias is negligible for all clusters with relative parameter uncertainties of $\lesssim10\%$ during the all sky survey ($t_{\text{exp}}=1.6\text{ ks}$). This is independent of the available information on the redshift. The same result is observed for $t_{\text{exp}}=20$ ks and for clusters with unknown redshift. Only clusters with available redshifts and precise temperatures in the deep exposure fields require a correction of the best-fit temperatures for distances above $\log(z)\gtrsim-0.5$, which is equivalent to $z\gtrsim0.32$.

\subsection{Bias in the error estimates}

For the bias in the error estimates no definite dependence on the input cluster mass or the redshift is observed, therefore, simple correction factors are calculated. Thus, we present estimates of these biases averaged over the complete simulated mass and redshift range (Tab. \ref{TabUncertainBias}). In analogy to the fit of the bias on the best-fit properties, masses with only catastrophic failures for all simulated redshifts are excluded from the estimation.\\\indent
\begin{table}
\centering
	\begin{tabular}{c|c|c|c}\hline\hline
		\multirow{2}{*}{simulation} & \multirow{2}{*}{bias} & \multicolumn{2}{c}{exposure time} \\
		 						  &							&  $t_{\text{exp}}=1.6$ ks & $t_{\text{exp}}=20$ ks\\\hline
		 known $z$			&	$\langle\Delta T_{\text{error}}\rangle/\Delta T$	 & 1 	& 1\\\hline
		 \multirow{2}{*}{unknown $z$}	& $\langle\Delta T_{\text{error}}\rangle/\Delta T$  & $\sim0.3$	& $\sim0.5$ \\
		 & $\langle\Delta z_{\text{error}}\rangle/\Delta z$ & $\sim0.25$ & $\sim0.15$\\\hline
	\end{tabular}
	\caption{Bias in the error estimates for the different simulations. The bias is avaraged over the complete mass and redshift range. For the simulations with known redshift, the bias in the uncertainties is in general negligible.}
	\label{TabUncertainBias}
\end{table}
If the redshift of the cluster is known, the temperature uncertainty computed by the {\tt{error}}-command well represents the statistical scatter in the best-fit values with a ratio in the uncertainties of $\langle\Delta T_{\text{error}}\rangle/\Delta T=1\pm0.1$. For spectral fits with unavailable redshifts, we observe a general underestimation of the proper uncertainty in the fit value by the {\tt{error}}-command, where the uncertainty in the redshift experiences a stronger bias than the uncertainty in the temperature (Tab. \ref{TabUncertainBias}).\\\indent
This increase in the bias for clusters with unknown redshift is explained by the additional free parameter during the spectral fit and by the difficulty in recovering the cluster redshift from X-ray spectra \citep{Yu2011}. Also, an increased exposure time does not necessarily result in a reduced bias in the error estimates. Unlike the bias in the best-fit parameter values, the bias in the error estimates needs to be considered commonly for the analysis of clusters with relative parameter uncertainties of $\lesssim10\%$. In the reduction of \textit{eROSITA} data for clusters with unavailable redshifts, the provided corrections are a necessary tool to compute reliable parameter uncertainties.


\section{Cosmological interpretation}
\label{Cosmo}

\begin{figure}[tb]
	\centering
	\includegraphics[scale=0.5, trim=0cm 0cm 0cm 1cm, clip=true]{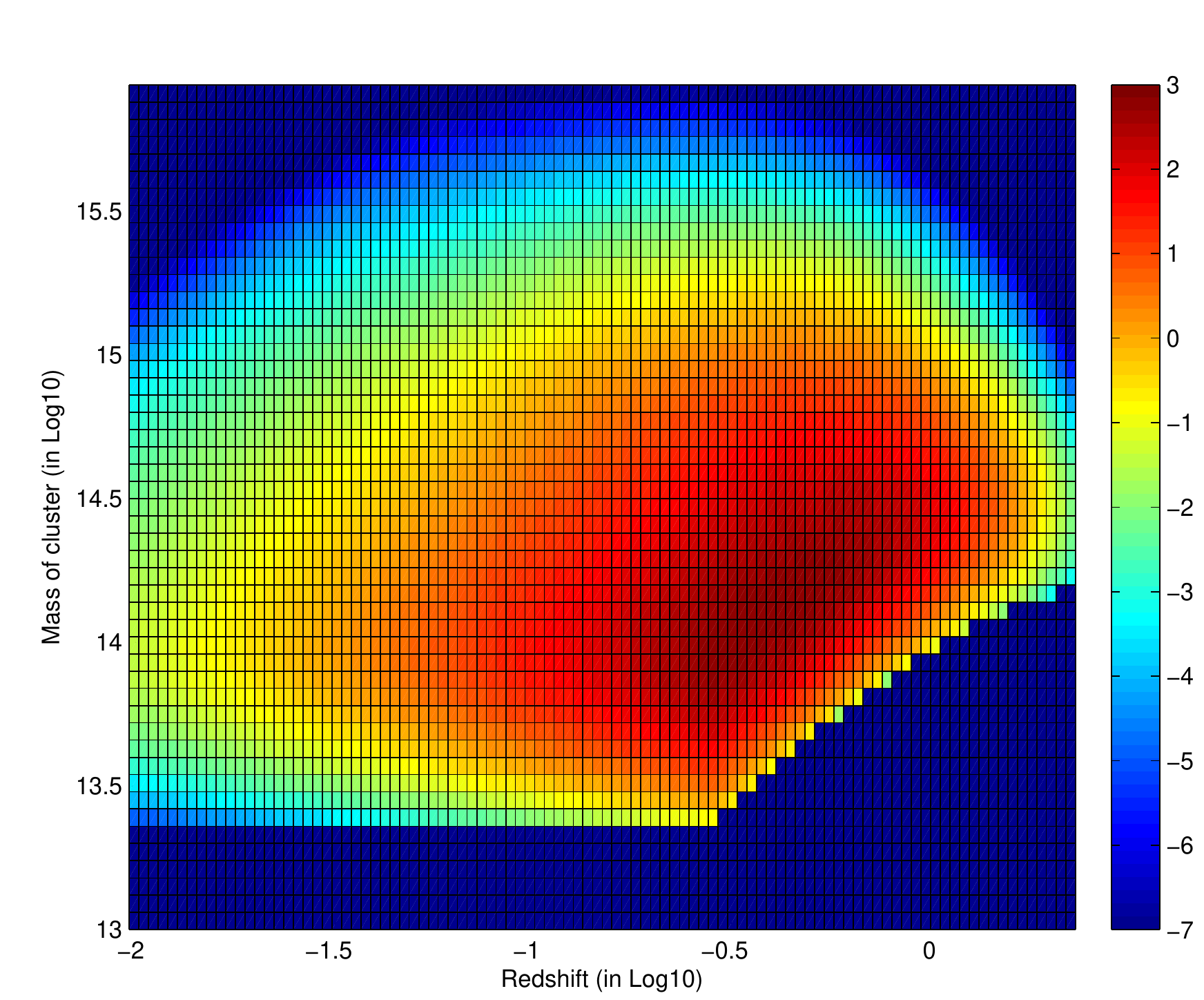}
	\caption{Distribution of galaxy clusters with mass and redshift as it will be detected by the \textit{eROSITA} instrument during its four years of all sky survey based on the mass function by \cite{Tinker2008} and on the scaling relations by \cite{Reichert2011}. The colour bar indicates the number of galaxy clusters in the individual bins in units of $\log_{10}$ and the cluster mass is considered in units of $\log(M/\text {M}_{\odot})$. We assume that a minimum number of $\eta_{\text{min}}=50$ photons is necessary to identify a cluster and effectively apply a lower mass cut to exclude low mass galaxy groups.}
	\label{PixDistribution}
\end{figure}

\begin{figure}[tb]
	\centering
	\includegraphics[scale=0.35,trim=0.5cm 1cm 0.5cm 2cm, clip=true]{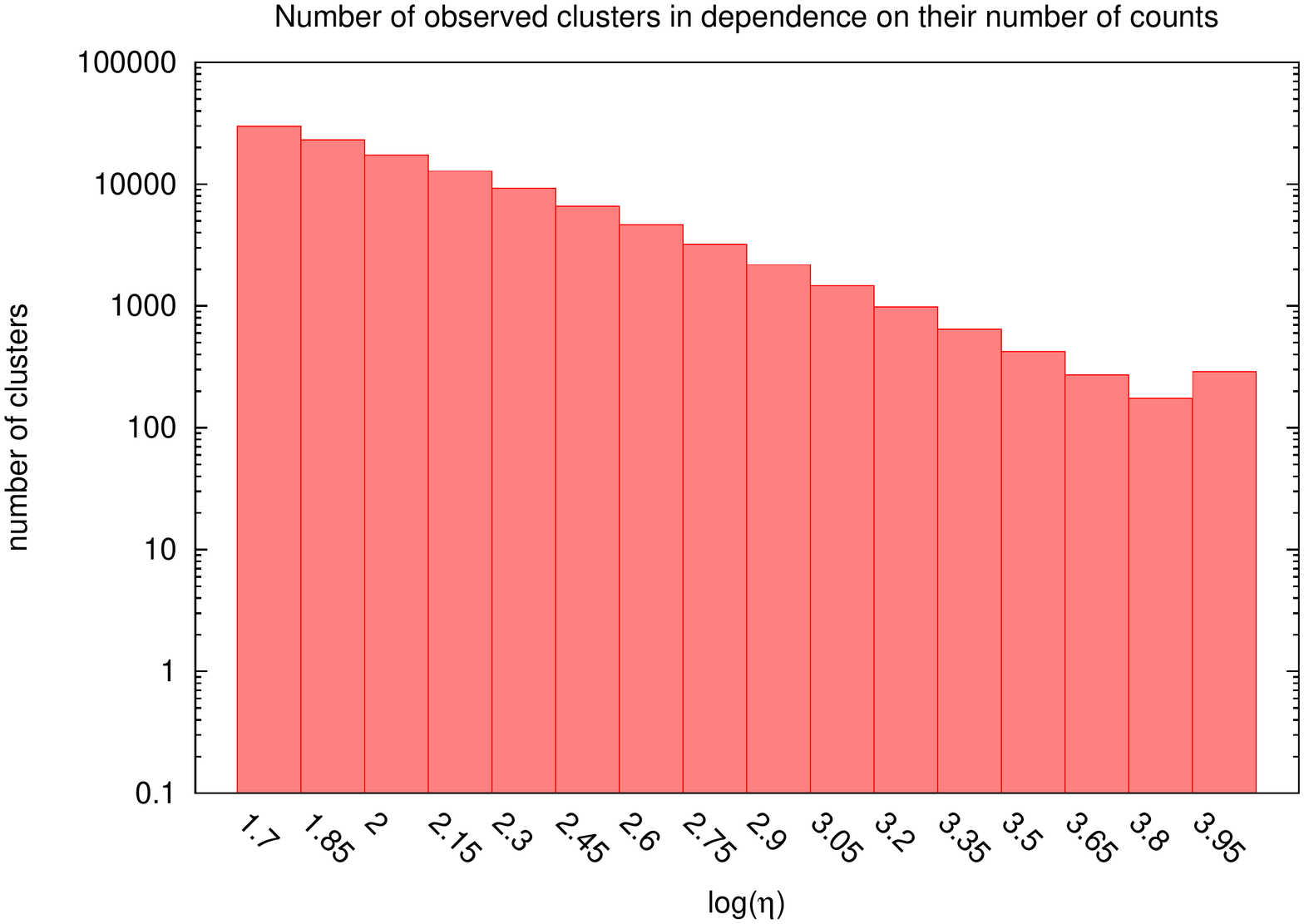}
	\caption{Number of observed galaxy clusters in dependence on their number of photon counts $\eta$ on logarithmic scale. The clusters are binned according to their number of observed photons in bins of the size $\Delta\log(\eta)=0.15$, starting at $\log(\eta)=1.7$ or $\eta=50$, respectively. The final bin includes all clusters with more than $\log(\eta)=3.95$ or $\eta\approx 9,000$ counts.}
	\label{PixHistogramm}
\end{figure}

In order to compute the number of clusters for which high precision temperatures and redshifts will be available directly from \textit{eROSITA} data, we apply the halo mass function by \cite{Tinker2008}. This mass function is in a first step convolved with the $M-L$ as well as with the $M-T$ scaling relation by \cite{Reichert2011} to yield a halo mass function in dependence on the input mass and redshift. In a second step, we apply the \textit{eROSITA} response on the above function and yield a distribution of the number of clusers in dependence on the number of observed photons. As in section \ref{Results}, the results are dependent on the input cluster properties. Figure \ref{PixDistribution} presents this distribution of clusters for an exposure time of $t_{\text{exp}}=1.6\text{ ks}$. For our computation we assume a minimum number of photons $\eta_{\text{min}}=50$ in the energy range of ($0.5-2.0$) keV for a source to be detected as a galaxy cluster by \textit{eROSITA} \citep[following][]{Pillepich2011}. Accordingly, no constant flux cut is applied for our computations, but for each considered combination of cluster mass and redshift the number of observed counts is estimated based on the applied scaling relations. Additonally, we apply an effective lower mass cut of $M_{\text{cut}}=5\times10^{13}/h$ M$_{\odot}$, which is equivalent to $M_{\text{cut}}=7.1\times10^{13}$ M$_{\odot}$ for our choice of $h=0.7$. With this latter cut we remove low mass clusters and groups, which show a strong scatter in their scaling relations \citep[e.g.][]{Eckmiller2011}. During the simulation, this mass cut is converted into a redshift dependent cut of the photon counts as explained by \cite{Pillepich2011}, since for the analysis of X-ray data the cluster mass is initially unknown. According to our applied cosmology ($\Omega_{\text{m}}=0.3$, $\Omega_{\Lambda}=0.7$), we adjust the normalisation of the matter power spectrum to $\sigma_8=0.795$ by means of the relation
\begin{equation}
\sigma_8\propto\Omega_{\text{m}}^{-0.38}
\label{eqSigma8}
\end{equation}
\citep{Reiprich2002}, which we normalise according to the WMAP5 results of $\Omega_{\text{m}}=0.279$ and $\sigma_{\text{8}}=0.817$ \citep{Komatsu2009}. This normalisation is chosen for a better comparison between our calculations and the work by \cite{Pillepich2011}. We define the observed sky fraction to be $f_{\text{sky}}=0.658$ for the all-sky survey with $t_{\text{exp}}=1.6$ ks. This sky fraction considers the entire sky, excluding a region of $\pm20^{\circ}$ around the Galactic plane as well as regions with a high X-ray flux such as e.g., the Magellanic Clouds and the Virgo Cluster.\\\indent
Following these approaches we expect to detect a total of \\$\sim113,400$ clusters of galaxies with the \textit{eROSITA} instrument during its four years of all-sky survey (Tab. \ref{TabNumbers}). The peak of the cluster distribution is located at a redshift of $\log(z)\approx-0.5$, $z\approx0.3$, and at a cluster mass of $\log(M/M_{\odot})\approx14$ \citep[compare][]{Pillepich2011}. For the highest cluster masses, the number of observed clusters is strongly limited at the local redshifts (Fig. \ref{PixDistribution}) due to the small observed volume. Also, at the highest redshifts we do not expect to detect any high mass clusters according to our concordance cosmology, which disfavours the existence of massive clusters at high redshifts. Galaxy clusters with low masses of $\log(M/M_{\odot})\lesssim14$ only show small fluxes at high redshifts of $\log(z)\gtrsim0.3$, which results in less than $50$ photons for an exposure time of $t_{\text{exp}}=1.6\text{ ks}$, and thus does not allow for a detection. Figure \ref{PixHistogramm} presents the distribution of the observed clusters in dependence on their number of photon counts for the all-sky survey. As a rough estimate, the currently known X-ray clusters are located in the two bins with the highest counts. Accordingly, this graphic emphasises the large amount of so far unknown clusters which will be discovered by \textit{eROSITA}. \\
For the analysis of the deep exposure fields with $t_{\text{exp}}=20$ ks, the sky coverage is re-defined to be $f_{\text{sky}}=0.0034$ \citep{Merloni2012}, such that the total number of observed clusters for these regions decreases to $2,600$. At the same time the clusters are observed up to more distant redshifts in these deep fields.\\\indent
In convolving this number distribution of \textit{eROSITA} clusters with the results obtained in Sec. \ref{RelUncertain}, we compute as a first estimate the number of clusters for which \textit{eROSITA} will detect precise temperatures and redshifts in addition to the already studied $184$ \textit{eHIFLUGCS} clusters (Tab. \ref{TabNumbers}). For this we integrate over the mass and redshift space with precise cluster properties, where we define the integration boundaries to be centred between the last pixel within this precise parameter space and its neighbouring pixel. Also, we investigate the compatibility between the assumed limit of $\eta_{\text{min}}=50$ for the detection of a cluster and the required limit of 100 counts for the reliable analysis of the cluster spectrum. Even though these two limits are based on different energy bands, $(0.5-2.0)$ keV and $(0.3-8.0)$ keV, respectively, all clusters, analysed in section \ref{Results}, are within the detection limit.  According to these assumption, \textit{eROSITA} is expected to obtain precise temperatures for $\sim1,670$ clusters during its all-sky survey if the redshift of the clusters is already known. This number of precision clusters emphasises the importance of this instrument as the number of clusters with precise temperatures will be increased by a factor of $\sim9$ compared to \textit{eHIFLUGCS}. Assuming the redshifts to be unavailable for all clusters, the number of clusters with precise temperatures decreases to $\sim300$ as the parameter space of precise temperatures reduces significantly (compare Fig. \ref{Pix16ksVarZRelT}). For all of these $300$ clusters precise X-ray redshifts will be available as well from \textit{eROSITA} data. Additionally, the simulations predict \textit{eROSITA} to obtain precise X-ray redshifts with relative uncertainties of $<10\%$ for a total of $23,000$ clusters. This entire cluster sample can then be employed for cosmological studies where a first estimate can already be obtained knowing only the cluster redshift and luminosity \citep[compare][]{Pillepich2011}. Following Table \ref{TabNumbers}, the percentage of \textit{eROSITA} clusters with precise properties increases significantly with increasing exposure time, which is allowing us an outlook also into the successive pointed observation phase of the mission. Only the redshift estimates in the deep exposure fields are significantly limited by catastrophic failures in the spectral fit.\\\indent
Even though we define a minimum number of photons of $\eta_{\text{min}}=50$ for a galaxy cluster to be detected by \textit{eROSITA}, the number of clusters with precise properties is limited by the $100$ photon counts, required for a reliable analysis of the cluster spectrum (Sec. \ref{SimOutline}). However, the application of $\eta_{\text{min}}=50$ for the computation of the number of clusters allows for a comparison of the number of clusters with precise property values to the total number of observed clusters. If we assume a less conservative approach with $\eta_{\text{min}}=100$, the total number of observed clusters in the all-sky survey decreases to $60,100$, whereas the number of clusters with precise properties remains the same. With this assumption, the percentages stated in Table \ref{TabNumbers} increase significantly, e.g. to $\sim2.8\%$ for clusters with known redshift in the all-sky survey.

\begin{table}[tb]
	\centering
	\begin{tabular}{cc|c|c}
		\hline\hline
		\multicolumn{2}{c|}{\multirow{2}{*}{simulation}} & $t_{\text{exp}}=1.6$ ks,  & $t_{\text{exp}}=20$ ks, \\
			&		& $f_{\text{sky}}=0.658$ &  $f_{\text{sky}}=0.0034$ \\\hline
			\multicolumn{2}{c|}{total}				  &	$113,400$	& $2,600$		\\\hline
		known $z$ & precise $T$				  &      $1,670$ ($\sim1.5\%$)      & $280$ ($\sim11\%$)          \\\hline
		\multirow{2}{*}{unknown $z$}	& precise $T$			  &	$300$ ($\sim0.3\%$)	&  $140$ ($\sim5\%$)		\\
			& precise $z$ & $23,000$ ($\sim18\%$)	&  $340$ ($\sim13\%$)		\\ \hline	 	
	\end{tabular}
	\caption{Number of clusters expected to be detected by \textit{eROSITA} in total, with relative temperature uncertainties of $\lesssim10\%$, when assuming the cluster redshift to be available, and with relative uncertainties of $\lesssim10\%$ in temperature and redshift in the case of unavailable redshift. The presented numbers for the precision clusters refer to clusters with fluxes of $F<9\times10^{-12}$ erg/s/cm$^2$, i.e. clusters without high quality observations already studied through \textit{eHIFLUGCS}. The values in parentheses denote the fraction of clusters with precise X-ray properties compared to the total number of clusters for each exposure time.}
	\label{TabNumbers}
\end{table}


\section{Discussion}
\label{Discussion}

\subsection{Dependence of the relative uncertainties}
\label{DependenceUncertain}

The fit of the model emission to the cluster spectrum is generally guided by the observed spectral lines, the over-all shape of the spectrum as well as by the position of the exponential cut-off at high energies. For clusters with temperatures of $\text{k}T\lesssim2.5$ keV, the fit is dominated by the line emission, as most emitted photons are observed in this spectral characteristic. As the cluster temperature increases, the spectral shape and the cut-off become more important for the fit.\\\indent
In section \ref{RelUncertain} as well as in Figures \ref{Pix16ksRelT} through \ref{Pix20ksVarZRelZ} we see a general increase of the relative uncertainties with increasing redshift and with decreasing cluster mass. This dependence is explained by the following aspects.\\\indent
For a constant cluster luminosity, the photon flux strongly declines with increasing redshift as $F\propto1/D_{\text{L}}^2$ with the luminosity distance $D_{\text{L}}$. This reduction is alliviated, but not fully compensated by the increase in luminosity with rising redshift, if we consider clusters with a constant mass (Eq. \ref{EqL-M}), such that the uncertainty of the fit parameters increases with increasing redshift. However, clusters with increasing total mass yield a strong increase in their luminosities, which improves the fit results despite the higher temperatures of these clusters. These increased temperatures result in a depletion of the emission lines as well as in a shift of the position of the exponential cut-off to higher energies and thus out of the \textit{eROSITA} effective area.\\\indent
The parameter space of clusters with precise properties extends to larger distances for the increased exposure time of $t_{\text{exp}}=20\text{ ks}$ as more photon counts are observed from the individual clusters and the statistical scatter in the spectrum is reduced. However, as already expressed in section \ref{RelUncertain}, the relative uncertainties are not only depending on the number of detected source photons, but also on the cluster characteristics. These characteristics include especially the strength of emission lines and the position of the high energy cut-off in comparison to the \textit{eROSITA} effective area.

\subsection{Further remarks on the relative uncertainties}
\label{FurtherUncertain}

According to our simulation results, we expect \textit{eROSITA} to detect X-ray redshifts for $\sim23,000$ clusters, which appears as an optimistic number at first glance. To test the reliability of these results, we analyse the relative redshift uncertainties for the two \textit{eHIFLUGCS} clusters RXCJ 1504 and A2204, kindly provided by G. Schellenberger. Both clusters show high redshifts of $z=0.215$, $\log(z)=-0.67$, and $z=0.15$, $\log(z)=-0.82$, with masses of $M=10^{15}\text{ M}_{\odot}$ and $M=7\times10^{14}\text{ M}_{\odot}$, $\log(M/\text{M}_{\odot})=14.85$, respectively. To allow for a comparison, the exposure times of the two \textit{Chandra} observations are decreased to $t_{\text{exp}}<2$ ks and only the temperature, the redshift and the normalisation of the spectrum are left free to vary during the fit. With this approach RXCJ 1504 and A2204 show a relative redshift uncertainty of $\Delta z/(1+z)\approx0.04$ and of $\Delta z/(1+z)\approx0.07$, respectively. Furthermore the best-fit redshifts very well represent the true redshifts with a deviation of only a few percent in the case of RXCJ 1504 and with no deviation for A2204. This result is well in accord with the precise redshift estimates for clusters with large distances, obtained in our simulations (compare Fig. \ref{Pix16ksVarZRelZ}). Furthermore, the analysed \textit{eHIFLUGCS} clusters are located in a parameter range, in which our simulations predict a large fraction of catastrophic failures (compare Fig. \ref{Pix16ksVarZRelZ}). According to this, the above analysis of observed data illustrates the conservative approach of our simulations to re-obtain the cluster properties.\\\indent
Since the estimation of ICM metallicities commonly presents large uncertainties when analysing observed data \citep{Balestra2007,Werner2008,Baldi2012}, we quantify the effect of a wrongly assumed metallicity on our simulations. As the metallicity presents itself especially in the strength of the emission lines, we only expect the metallicity to influence our results for clusters with $\text{k}T\lesssim2.5$ keV. To test this influence, we repeat our simulation for a choice of clusters with different masses and redshifts, where the cluster temperature meets the above criterion and the redshift is assumed to be known. During the fitting procedure, the metallicity is wrongly fixed to the extreme values of either $A=0.2\text{ A}_{\odot}$ or $A=0.4\text{ A}_{\odot}$ instead of the true value $A=0.3\text{ A}_{\odot}$ \citep{Maughan2008}. Even for these strong deviations in the metallicities, the relative temperature uncertainties only display an increase for the more distant clusters of  $\log(z)\gtrsim1.1$, $z\gtrsim0.1$, by a couple of percent. However, the accuracy of the temperature fit is unaffected by the wrongly fixed metallicity. \\
Since the metallicity of a cluster is not only definded by the value of $A$, but also by the applied abundance model, we repeat our simulation for $t_{\text{exp}}=1.6$ ks and a sample of clusters with the more recent abundance model by \cite{Asplund2009}. Assuming the redshift of the tested clusters to be known, we obtain differences in the relative temperature uncertainties of $\sim5\%$ and differences of only a couple of percent for the bias in the temperature estimates. These differences do not show an apparent dependence on the simulated cluster properties. In summary, we conclude that neither a wrongly fixed metallicity nor a change in the abundance model alters the simulated parameter spaces or the numbers of clusters with precise properties significantly.\\
A possible evolution of the metallicity with redshift could not be definitely quantified, yet, and we thus apply a constant metallicity in our simulations (compare Sec. \ref{ClusterProp}). Assuming a metallicity evolution would impact our simulation results for the higher redshifts as the metallicity might decrease to half its value at redshifts of $z\approx1$ \citep[e.g.][]{Maughan2008}. Since redshift estimates will be possible up to $z\approx0.3$ (compare Fig. \ref{Pix16ksVarZRelZ}) in the all-sky survey, we quantify the influence of such an evolution on the redshift analysis. In an extreme scenario of $A=0.2$ A$_{\odot}$ for clusters at $z\approx0.3$, the relative redshift uncertainties increase to $\lesssim12\%$. This results in a shift of the contour line of $\Delta z/(1+z)<10\%$ to lower redshifts by one pixel. However, the tested scenario requires an unanticipatetly strong metallicity evolution with an already strong decrease in metallicity over a small redshift range in contrast to the literature \citep[e.g.][]{Balestra2007,Maughan2008}.\\
Despite our realistic treatment of the background and its statistical scatter (compare Sec. \ref{background}), systematic errors in the anaysis of future observed data might arise due to a wrongly constrained background model. To investigate its effect on our results, we rerun the simulations for a set of parameters of typically observed \textit{eROSITA} cluster masses of $M\approx10^{14}$ M$_{\odot}$ up to $M\approx10^{14.8}$ M$_{\odot}$ with relative temperature uncertainties of $\sim10\%$. In these simulations we then assume a background model with a systematic error of $\pm10\%$. This is a conservative approach given that e.g. for \textit{Chandra} uncertainties of $\sim3\%$ are quoted \citep{Markevitch2003}, such that we are expecting a reduced value for \textit{eROSITA}. For clusters with precise parameter estimates and low temperatures of ${\rm{k}}T\lesssim3$ keV ($M\lesssim10^{14}$ M$_{\odot}$), the difference in the newly simulated parameter bias and in the relative uncertainty is only around a couple of percent when compared to the simulation without applying any background uncertainty. These differences slightly increase to $\sim10\%$ for clusters with precise parameters, but intermediate temperatures, corresponding to $M\approx10^{14.8}$ M$_{\odot}$. This is true for the all-sky survey as well as for the deep exposures. According to this, introducing a possible background error in our simulations does not influence the presented parameter space of clusters with precise properties. Also, the temperature bias still remains negligible for clusters within this parameter space, apart from the exclusions already stated in section \ref{ParamBias} for clusters observed in the deep exposure fields.

\subsection{Occurence of catastrophic failures}
\label{Failures}

As described in section \ref{RelUncertain}, catastrophic failures particularly occur for spectral fits with unknown cluster redshifts, especially for the very low mass and the very high mass clusters (e.g. Fig \ref{Pix16ksVarZRelT}). This finding is generally explained by the degeneracy between the redshift and the temperature for these cluster masses.\\
This degeneracy in dependence on the cluster mass and temperature is illustrated in figure \ref{PixErrorEllipse}, where we plot the distribution of temperature and redshift best-fit values for three different parameter sets each with roughly the same number of counts. The low and the high mass parameter set is rejected due to large numbers of catastrophic failures and both sets show a strong correlation between their best-fit redshifts and temperatures. The stripe features, especially visible in the top image, are the result of the {\tt{steppar}}-fit and are addressed in Sec. \ref{DisStrategy}. \\
To explain this degeneracy and the simulation results for the clusters with unknown redshift several spectral charactersitics interplay with one another. We find two possible examples for the explanation of the simulation results in the strength of the emission lines, especially in the strength of the Fe-K line, and in the detectability of the exponential cut-off.\\
Low mass clusters only show small numbers of detected photons and thus a large statistical scatter in their spectra. Additionally, the individual emission lines are not resolved (compare Fig. \ref{PixClusterSpectrum}) and the observed emission line complexes around energies of $1$ keV are shifting to higher energies with increasing temperatures. This latter characteristic thus leads to a degeneracy between the imprint of the redshift and the temperature on the spectrum. Furthermore, due to the scatter in the emission lines at the energies of the exponential cut-off, the exact energy of this spectral feature is not detectable, which complicates the spectral fits. Considering these two aspects, we explain the large fraction of catastrophic failures for the fit to spectra of clusters with low masses (compare e.g., Fig. \ref{Pix16ksVarZRelT}). Only for higher cluster temperatures of $\text{k}T\gtrsim2.5\text{ keV}$ and thus with fading emission lines, this degeneracy is partially lifted. For these clusters the spectral fit is mainly guided by the position of the exponential cut-off, which is no longer obscured by the emission lines, and by the Fe-K line, which increases in strength with increasing cluster temperatures. However, if we consider clusters with even higher masses of $M\gtrsim10^{15}$ M$_{\odot}$ as well as higher redshifts, which of the two competing effects, higher temperatures or higher redshifts, dominates the shift of the exponential cut-off? \\
To answer this question, we investigate the position of the exponential cut-off in dependence on the cluster mass and redshift for clusters with roughly the same number of source counts. As displayed in figures \ref{Pix16ksRelT} - \ref{Pix20ksVarZRelZ}, the contour lines of constant counts can be approximated as linear functions with a slope of $m=1$. Moving up along this contour line, both cluster mass and redshift increase by a factor of $\Delta=10^{0.15}=1.41$ with every pixel (compare Sec. \ref{ClusterProp}).
According to the emissivity of the thermal bremsstrahlung
\begin{equation}
	\epsilon_{\nu}^{\text{ff}}\propto T^{-1/2}\cdot\text{e}^{-\frac{h\nu}{\text{k}T}}\quad ,
\end{equation}
the position $E$ of the exponential cut-off, where $\epsilon_{\nu}^{\text{ff}}\propto 1/\text{e}$, is proportional to the cluster temperature. When also considering the cluster redshift, the position of the cut-off shows the relation
\begin{equation}
	E\propto \frac{T}{(1+z)}\quad ,
\end{equation}
such that the ratio between $E_1$ and $E_2$ for two neighbouring pixels along the line of constant photon counts derives as
\begin{equation}
	\frac{E_2}{E_1}=\frac{T_2}{T_1}\cdot\frac{(1+z_1)}{(1+z_2)}=\frac{T_2}{T_1}\cdot\frac{(1+z_1)}{(1+z_1\cdot\Delta)}\quad ,\label{EqCutOffRatio}
\end{equation}
with $z_2>z_1$. According to equation \ref{EqT-M} the ratio between the two temperatures is defined by the $M-T$ scaling relation
\begin{eqnarray}
	\frac{T_2}{T_1}&=&\left(\frac{M_2}{M_1}\right)^{0.62}\cdot\left(\frac{\Omega_{\text{m}}\cdot(1+z_2)^3+\Omega_{\Lambda}}{\Omega_{\text{m}}\cdot(1+z_1)^3+\Omega_{\Lambda}}\right)^{0.32}\\
	&=&\Delta^{0.62}\cdot\left(\frac{\Omega_{\text{m}}\cdot(1+z_1\cdot\Delta)^3+\Omega_{\Lambda}}{\Omega_{\text{m}}\cdot(1+z_1)^3+\Omega_{\Lambda}}\right)^{0.32}\quad .\label{EqTRatio}
\end{eqnarray}
Combining expressions \ref{EqCutOffRatio} and \ref{EqTRatio}, we obtain the final ratio of the position of the exponential cut-offs along the line of constant photon counts
\begin{equation}
	\frac{E_2}{E_1}=\Delta^{0.62}\cdot\left(\frac{\Omega_{\text{m}}\cdot(1+z_1\cdot\Delta)^3+\Omega_{\Lambda}}{\Omega_{\text{m}}\cdot(1+z_1)^3+\Omega_{\Lambda}}\right)^{0.32}\cdot\frac{(1+z_1)}{(1+z_1\cdot\Delta)}\quad .
\end{equation}
A graphical analysis of this function indicates a ratio of $\frac{E_2}{E_1}>1$ for our choice of $\Delta=1.41$ and for the entire simulated redshift range. This result emphasises the shift of the exponential cut-off to higher energies for clusters with increased masses and redshifts along the lines of constant photon counts. In fact, for all $\Delta>1$ the result of $\frac{E_2}{E_1}>1$ holds true. For clusters with masses of $\log(M/\text{M}_{\odot})\gtrsim15$ for which catastrophic failures occur in the simulation with unavailable redshift the exponential cut-off is located at energies of $E\gtrsim8$ keV and thus out of the spectral fitting range. The thus arising difficulty in the spectral fit is additionally appended by the decreasing S/N-ratio for clusters with the same number of source photons, but with increasing redshifts. This evolution of the S/N-ratio with increasing redshift is explained by the increasing extend of the cluster, from e.g. $R_{\text{500}}\approx5.7$ Mpc for a cluster with $\sim1000$ counts at $z\approx0.08$ to $R_{\text{500}}\approx8.8$ Mpc at $z\approx0.45$, and the thus rising background emission.

\subsection{Influence of the analysis strategy}
\label{DisStrategy}

To test the reliability of our predictions, we analyse the influence of the simulation setup on our results.\\\indent
For several parameter sets we re-run the simulation with $500$, $700$ and $1000$ repetitions and compare the outcome to the results above for 300 repetitions. The changes in the biases and in the relative uncertainties for both temperature and redshift are only a few percent and these deviations become negligible for clusters with relative uncertainties of $\lesssim0.1$. An equivalent development is observed when altering the number of steps within the {\tt{steppar}}-fit. Even though with varying numbers of fitting steps the results show deviations of up to $20\%$ for parameter sets with high relative uncertainties, the results for clusters with high precise properties are comparable. We thus conclude that the parameter space of cosmologically interesting clusters with relative uncertainties of $\lesssim0.1$ in their properties is independent of the number of repetitions and of the number of steps in the more dimensional {\tt{steppar}}-fit. For these clusters already $300$ repetitions for the realisation of each parameter set yield a proper statistical, Gaussian-like distribution of the fit results.\\\indent
However, a small bias might arise in the {\tt{steppar}}-fit with variable redshift for clusters with local redshifts of roughly $z\lesssim0.1$. This bias is observed for clusters with unknown redshift and with low masses of $\log(M/\text{M}_{\odot})\lesssim14$ as well as for 
\begin{figure}[H]
	\begin{flushleft}
	\includegraphics[scale=0.35,trim=0.5cm 1cm 1.0cm 2cm, clip=true]{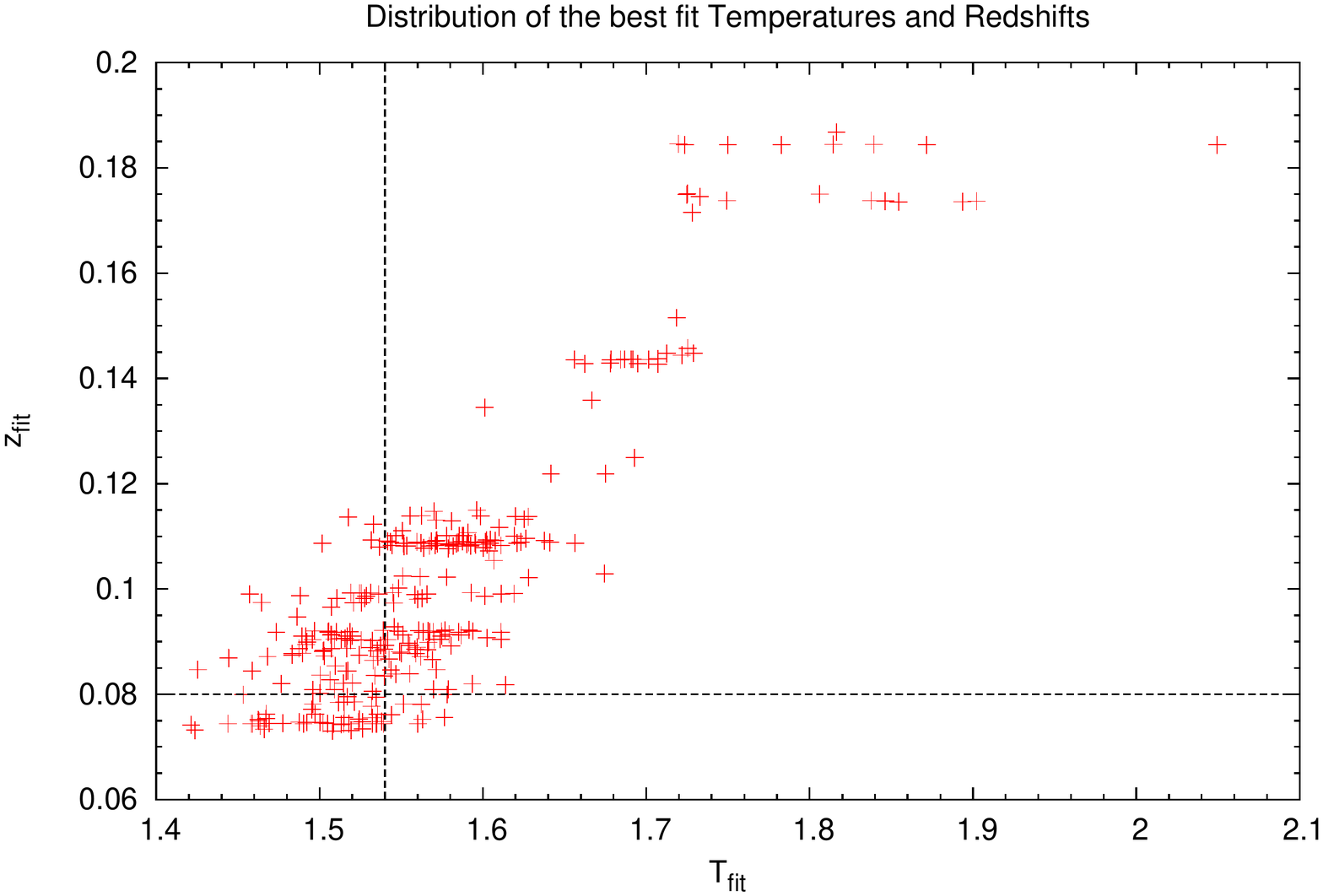}
	\includegraphics[scale=0.35,trim=0.5cm 1cm 1.0cm 2cm, clip=true]{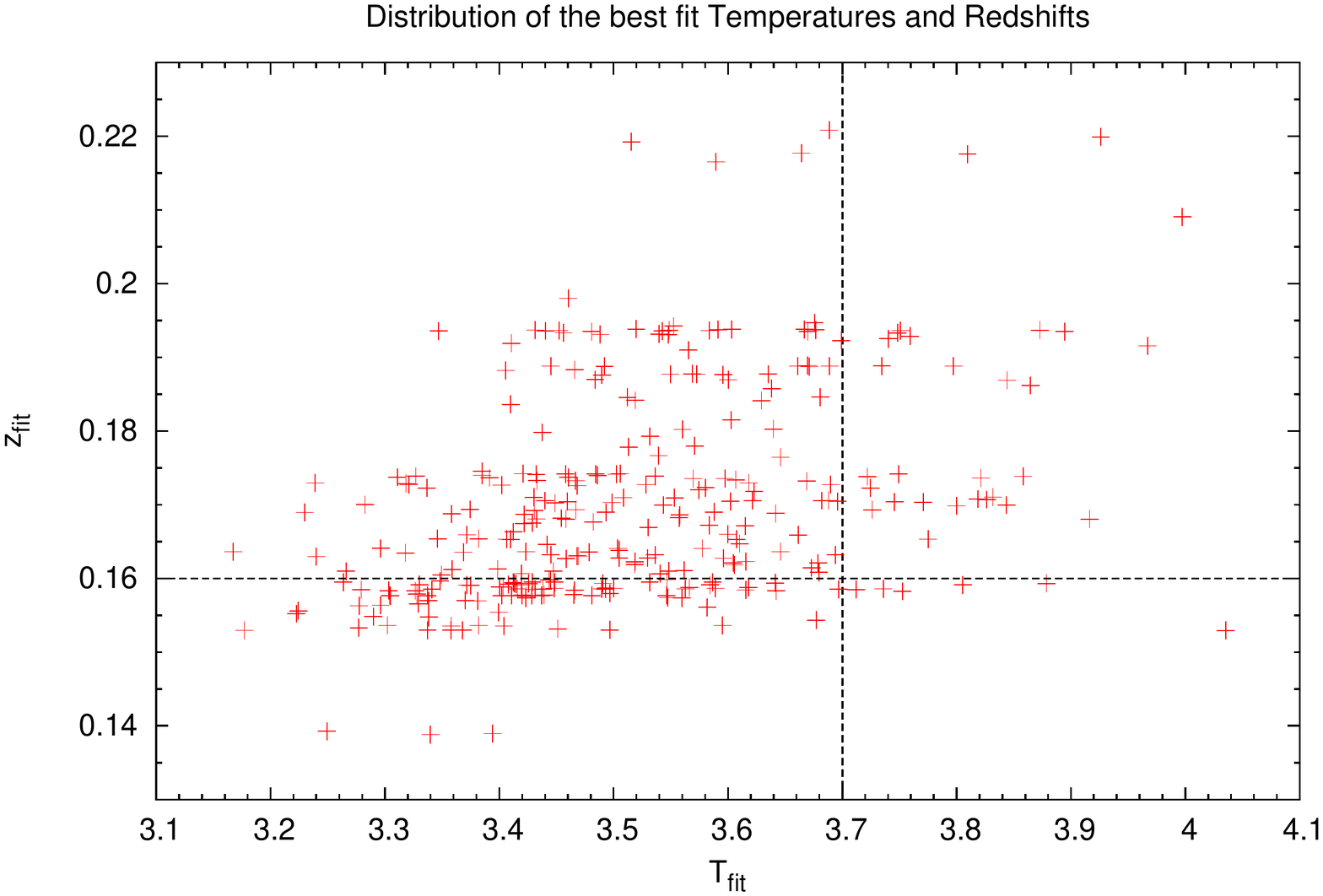}
	\includegraphics[scale=0.35,trim=0.5cm 1cm 1.0cm 2cm, clip=true]{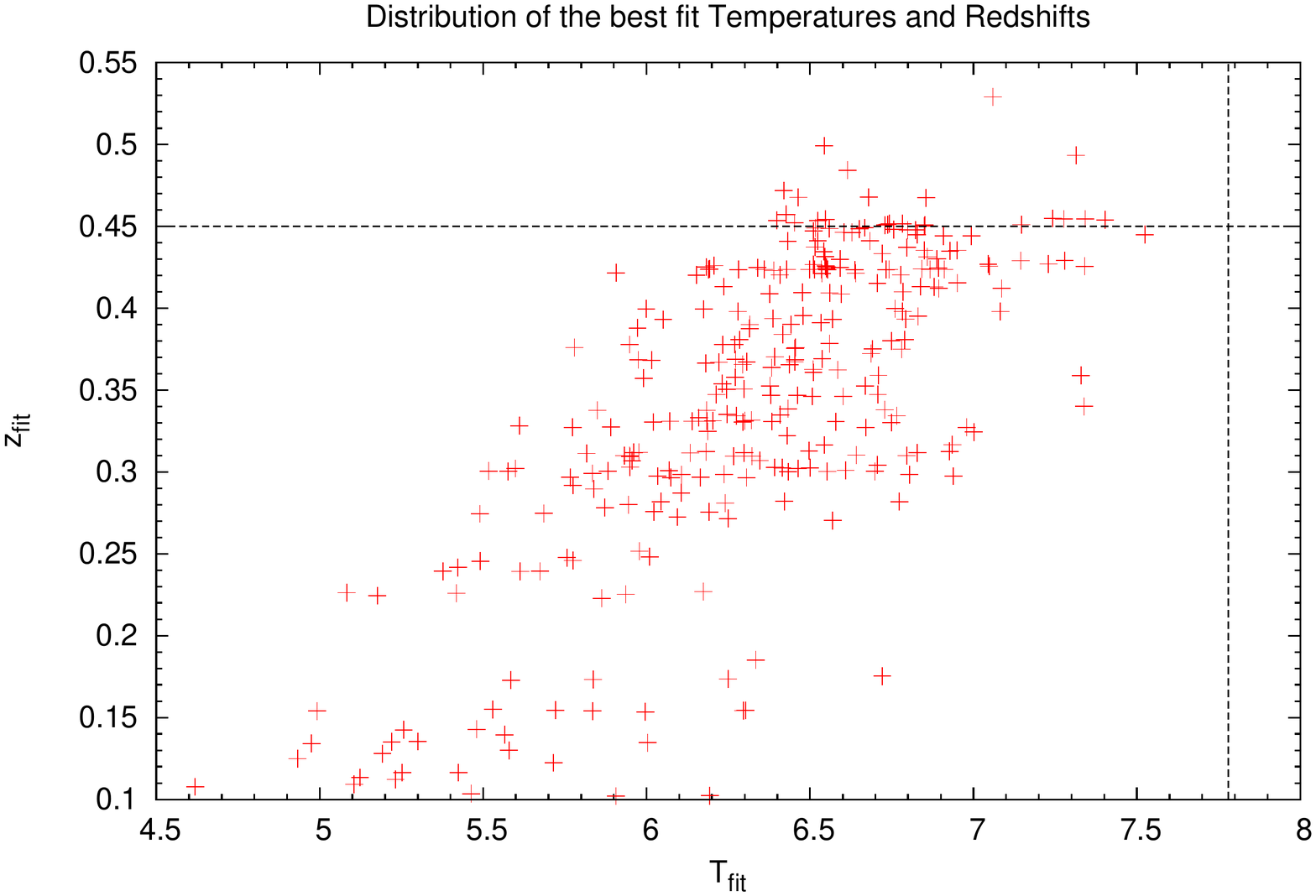}
	\end{flushleft}
	\caption{Distribution of the best-fit temperatures and redshifts for three different clusters in the deep exposure fields, each with roughly $5,000$ counts, but with different cluster masses and temperatures. From top to bottom: $\log(M/\text{M}_{\odot})=13.75$ $\log(M/\text{M}_{\odot})=14.35$, $\log(M/\text{M}_{\odot})=14.8$. The horizontal and vertical lines indicate the input redshifts and temperatures, respectively.  The low and high mass parameter set is rejected from the analysis due to large numbers of catastrophic failures. For these cluster masses, the correlation between the fit values of the temperature and the redshift emphasises the degeneracy between these two properties.}
	\label{PixErrorEllipse}
\end{figure} \newpage
intermediate mass clusters in the deep exposure fields (compare Figs. \ref{Pix16ksVarZRelZ} and \ref{Pix20ksVarZRelZ}). For these clusters statistical artifacts might arise (compare Fig. \ref{PixErrorEllipse} Top) since too little information is available for the fit.
We ran a thorough investigation for the fitting statistics of these clusters and conclude that cluster with artifact features are generally rejected due to large numbers of catastrophic failures. Even though, these clusters show a strong deviation between their input redshift and the starting value for the fit with $z=0.3$, the fit is not improved by an adaptation of the starting value.\\\indent
Generally, the simulated precisions and accuracies are not necessarily influenced by the starting values of the spectral fit, so that we apply commonly observed values of $\text{k}T=2$ keV and $z=0.3$ for the start of the fit. Only in the simulations with unknown cluster redshift, the number of rejected data sets for both intermediate and high mass clusters (e.g. Fig. \ref{Pix20ksVarZRelZ}) at their highest simulated redshifts can be improved if we choose values close to the input parameter values for the start of the fit. In this case, a strong decrease in the biases as well as in the relative uncertainties of up to $\lesssim25\%$ of the former value is observed. This results in less catastrophic failures in the mentioned mass ranges and for the deep exposures the parameter space of clusters with high precision properties increases to higher redshifts. However, with this adaptation of the fitting strategy, the percentage of precision clusters changes only for the deep exposure fields and only by $<1\%$. According to this, our setup, which does not require any knowledge on the input properties, presents a reliable estimate of the number of detected clusters with precise characteristics. \\\indent 
For the improvement of the analysis of future \textit{eROSITA} clusters with unavailable redshift, we suggest to re-fit the spectrum for different starting redshifts, where the starting value of the temperature is adapted to the redshift via an $L-T$ scaling relation \citep[compare][]{Lloyd2011}. The fit which returns the smallest parameter uncertainties is expected to as well record the greatest parameter accuracy.\\\indent
Finally, we also test the influence of the definition for rejected pixels on our results, since we require a minimum of $80\%$ of the repetitions to yield consistent and non-catastrophic data. This percentage emphasises that more than $20\%$ of unreliable fit results is unacceptable. For an increased minimum percentage of accepted data to $90\%$, the simulation results for $t_{\text{exp}}=1.6$ ks and clusters with known redshift remain unchanged. Within the other simulations, more parameter sets with especially high redshifts are rejected, in particular in the simulations with unknown redshift. However, this development reduces the parameter space of clusters with known redshift and relative temperature uncertainties of $\lesssim10\%$ only insignificantly.

\subsection{Remarks on the cosmological interpretation}
\label{RemarksNumbers}

Our simulations present an overview of the number of clusters for which \textit{eROSITA} will be able to obtain precise data. However, the future data reduction likely requires individual models for each observed cluster, which e.g. include individual background emissions and might thus slightly alter the presented numbers of clusters. In the previous sections of our discussion, we already concluded these numbers to be only insignificantly influenced by a wrongly assumed metallicity of the cluster and by the background emission.  For a possible evolution of the metallicity with redshift, however, the number of high and intermediate redshift clusters with precise properties might decrease depending on the scale of the evolution.\\
Additionally, the emission of a possible central AGN in clusters needs to be considered in the analysis of observed data, especially for deep exposures. In these observations, bright central AGN can impede the extraction of cluster spectra and even the detection of the clusters as extended source. Currently, investigations on the efficiency of different source detection algorithms are conducted. Meanwhile, we take another look at the simulation results for the deep exposure fields, which indicated the temperatures to be available with high precision up to the highest redshifts for high mass clusters (Fig. \ref{Pix20ksRelT}). Considering the above mentioned AGN confusion, a detection of precise temperatures up to a redshift limit of $z\approx1$ presents a more reliable and conservative estimate for those high mass clusters. With this redshift limit, however, the total number of precise clusters in the deep exposure fields remains uneffected since only very few clusters with the highest masses of $M\gtrsim10^{15}$ M$_{\odot}$ are found at $z\gtrsim1.0$ (compare Fig. \ref{PixDistribution}), in particular when limited to the one hundred square degrees for the deep exposure fields. The estimation of precise redshifts in these fields is already limited to $\log(z)\lesssim -0.35$, $z\lesssim0.45$, due to catastrophic failures (compare Fig. \ref{Pix20ksVarZRelT}) and is thus not influenced by AGN confusion.\\
\\
Recent simulations have shown the possibility of cosmological estimates with only luminosity and redshift information of the galaxy clusters available \citep{Pillepich2011}. Redshift information on \textit{eROSITA} clusters will be obtained through optical follow-up observations shortly after the launch of the mission. This work now presents for how many clusters also precise temperatures will be observed. In an up-coming work we will qualitatively test the improvement in the cosmological uncertainties with the help of these additional information. The cosmological analysis of cluster data is especially sensitive to the information coming from massive clusters. Our simulations now indicate that at the beginning of the \textit{eROSITA} survey precise information on massive clusters are rather difficult to obtain (compare Sec. \ref{RelUncertain} $\&$ \ref{Failures}). X-ray follow-up observations with eROSITA as well as with other instruments, such as e.g. \textit{XMM} or \textit{Astro-H}, will soon after determine the surface brightness and the temperatures of massive clusters. These information will then put tighter constraints on the cosmology, even though not for all of the massive clusters temperature estimates will be accessible, due to the large numbers of observed clusters.

\subsection{Comparison between different scaling relations}
\label{CompScaling}

\begin{figure}[tb]
	\centering
	\includegraphics[scale=0.35,trim=0.5cm 2cm 0.0cm 2cm]{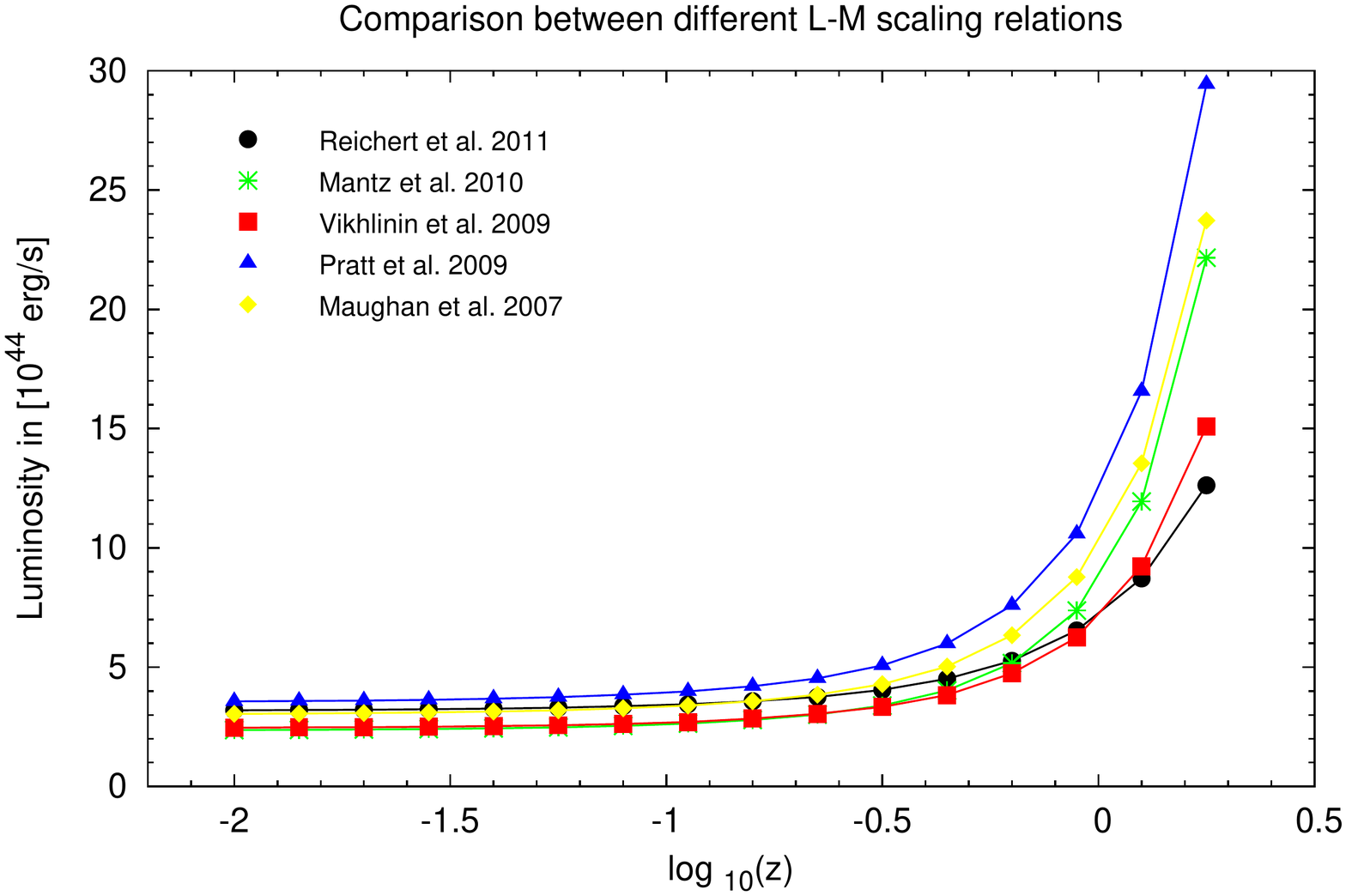}
	\caption{Luminosity in dependence on the cluster redshift for different scaling relations and a cluster mass of $M\approx5\times10^{14}$ M$_{\odot}$. The luminosities are computed within the energy range of ($0.1-2.4$) keV for all relations.}
	\label{PixLumi}
\end{figure}

We compare five commonly applied scaling relations \citep{Maughan2007,Pratt2009,Vikhlinin2009,Mantz2010,Reichert2011} with one another and analyse the effects of a change in the scaling relation on the results of our simulations. For a recent review on cluster scaling relations see \cite{Giodini2013}.\\\indent
The five different $M-T$ relations deviate from one another especially for the smallest cluster masses of $\log(M/\text{M}_{\odot})\lesssim14$ with an increasing inconsistency for increasing redshifts. The scaling relation by \cite{Mantz2010} shows the strongest increase of the temperature with the cluster mass for a fixed redshift and the relation by \cite{Maughan2007} presents the shallowest slope. The work by \cite{Reichert2011} approximates an average value for the slope. The luminosities computed by means of the different considered scaling relations for a fixed cluster mass are very comparable at the local redshifts (Fig. \ref{PixLumi}). For a cluster mass of $\log(M/\text{M}_{\odot})=14$ they start to deviate from one another for $\log{z}\gtrsim-0.5$, $z\gtrsim0.3$, where this deviation starts at lower redshifts for declining cluster masses. Within this comparison, the $M-L_{\text{X}}$ relation by \cite{Reichert2011} exhibits the most moderate evolution of the luminosity with redshift. The shallow development of the $L-M$ relation with redshift by \cite{Reichert2011} favours the application of this scaling relation, as distant clusters with $z\gtrsim0.3$ show smaller luminosities than the other scaling relations and thus fewer source counts. This characteristic is especially important for the simulation of the deep exposure fields, in case the cluster redshift is available. In the remaining three simulations, the parameter space of precise cluster porperties is mainly located at lower redshifts for which all considered scaling relations are comparable.\\\indent
The galaxy cluster sample, on which \cite{Reichert2011} base their findings, covers the largest mass and redshift range with $M=(5\times10^{13}-3\times10^{15})$ M$_{\odot}$ and $z\leqslant 1.46$, respectively, such that we only require a small extrapolation of this scaling relation to cover our simulated mass and redshift range. According to this aspect and to the evolution of the relations, the scaling relations by \cite{Reichert2011} describes the most conservative approach in terms of the characterisation of high-z clusters. \\\indent
The deviations in the individual scaling relations also result in differences in the distribution of clusters with mass and redshift (Appendix \ref{NumberComparison}). For example, due to the slightly lower luminosity in the scaling relation by \cite{Vikhlinin2009} at the local redshifts, the total number of clusters decreases to $\sim103,700$ compared to $\sim113,400$ clusters for the relations by \cite{Reichert2011}, when applying the same cosmology for both relations. However, the number of clusters with precise properties from \textit{eROSITA} data is comparable for both scaling relations with a deviation of $<2\%$ between the two. For example, for the scaling relation by \cite{Vikhlinin2009}, this deviation results in $1,700$ clusters with precise temperatures and already known redshifts for the all sky survey compared to the $1,670$ clusters for the relation by \cite{Reichert2011}.

\subsection{Comparison to other works}

Similar to the findings by e.g. \cite{Lloyd2011}, \cite{Planck2011} and \cite{Yu2011} our simulations depict cluster X-ray spectra as sensitive estimators of the redshift of the object. However, our simulations forecast the determination of cluster redshifts for the \textit{eROSITA} instrument as well as for exposure times as low as $t_{\text{exp}}=1.6$ ks for the first time. Our findings for the \textit{eROSITA} deep fields are comparable to the work by the \cite{Planck2011}, who yield precise redshifts up to distances of $z=0.54$ for $t_{\text{exp}}=10$ ks with the XMM-Newton instrument, when not correcting for the underestimation of the uncertainties. Also, as well as the above mentioned publications, our work shows a decrease of the fit accuracy for the analysis of cluster spectra with unavailable redshift. \\\indent
We emphasise that the precision and the accuracy of the cluster properties are rather dependent upon the values of these cluster properties themselves and not only on the number of detected photons, equivalently to the analysis by \cite{Yu2011}. However, in contrast to their findings, our simulations predict X-ray redshifts to be available also for clusters with less than $1,000$ photon counts, if these clusters show temperatures of $\text{k}T\lesssim5.5$ keV (Figs. \ref{Pix16ksVarZRelZ} $\&$ \ref{Pix20ksVarZRelZ}). This aspect is explained by the difference in the instrumental spectral responses between \textit{eROSITA} and \textit{Chandra}, on whose data \cite{Yu2011} base their analysis. For two clusters with the same total number of detected photons, \textit{eROSITA} will show more photons in the soft energy band, which improves the fitting statistics especially for the low temperature clusters above.\\\indent
Our expected number of $\sim113,400$ \textit{eROSITA} clusters is increased by $\sim15\%$ compared to the analysis by \cite{Pillepich2011}, as this work applies the scaling relation by \cite{Vikhlinin2009} and the cosmological model of the WMAP5 results \citep[compare Sec. \ref{CompScaling} $\&$ \ref{NumberComparison}]{Komatsu2009}, instead of the scaling relation by \cite{Reichert2011} used in our calculations. However, if we base the computation on the same set-up as \cite{Pillepich2011}, we obtain a negligible deviation of only $1\%$ from their results, which emphasises on the reliability of our code.


\section{Summary and conclusions}

The upcoming \textit{eROSITA} mission presents a powerful tool to test our current cosmological model and especially to study the nature of dark energy by investigating the distribution of galaxy clusters with mass and redshift. Moreover, it will allow for the study of cluster physics, e.g. in terms of scaling relations, in unprecedented detail. With the simulations presented in this work, we predict the accuracy and the precision with which the \textit{eROSITA} instrument will be able to determine the cluster temperature and redshift, and we introduce the number of clusters for which these properties will be available.\\\indent
The highest precision and accuracy of the temperature and redshift are obtained for clusters at the most local redshifts. In general, the precision and the accuracy of the cluster properties do not only show a dependence on the number of detected photons, but as well on the cluster properties themselves, especially on the redshift. For the average exposure time during the \textit{eROSITA} all-sky survey, high precision temperatures will be available for clusters as distant as $z\lesssim0.16$ and the instrument will allow for precise X-ray redshifts up to $z\lesssim0.45$, where for the very local clusters the uncertainty in the redshift is even comparable to optical photometric estimates. However, for the simulation with unknown cluster redshifts, catastrophic failures occur within the spectral fit and limit the parameter space of high precision properties especially for the lowest and the highest masses $\log(M/\text{M}_{\odot})\lesssim14$ and $\log(M/\text{M}_{\odot})\gtrsim15$, respectively. These failures arise due to the redshift as additional free parameter in the fit and because of the thus resulting degeneracy between the redshift and the temperature. As \textit{eROSITA} cluster spectra prove as sensitive estimators of the redshift for local clusters with intermediate masses, optical follow-up observations are most effective, if they first cover clusters without reliable X-ray redshifts and we predict they will preferentially be found above $z\approx0.45$. Additionally, these follow-up observations will eventually allow for more precise redshift estimates also for clusters at lower redshifts.\\\indent
Within the \textit{eROSITA} deep exposure fields, X-ray redshift and temperature information will be stronger limited by catastrophic failures than for the lower exposure time. According to this, precise X-ray redshifts are only observed to the same maximum distance as for the all sky survey. In this aspect, our simulations follow the conservative approach of no constraints on the starting values of the fit. However, the number of catastrophic failures for the spectral fit of intermediate and high mass clusters can be reduced, if additional information on the starting values, e.g. through the coupling of the fit parameters by the $L-T$ relation or with the help of first redshift estimates from shallow optical surveys, are available.\\\indent
If the redshift of the clusters in the deep exposure fields is known, the percentage of clusters with precise temperatures still increases significantly to the highest redshifts. Even though these deep fields only cover a small sky fraction, the findings for these regions shed light on the expectations for the succeeding pointed observation phase. \\\indent
The entire parameter space of clusters with precise properties displays great parameter accuracies, such that for those clusters no parameter bias needs to be corrected for. Only for the long exposure times of $t_{\text{exp}}=20$ ks the bias in the temperature needs to be considered for clusters with available redshifts at distances of $z\gtrsim0.32$. We additionally introduce correction functions, which need to be applied to spectral fits of clusters with a bias in their best-fit properties. For the analysis of observed \textit{eROSITA} data, these correction functions should be applied iteratively. The analysis of spectral cluster data yields preliminary values of the cluster temperature, redshift and luminosity from which the total mass is estimated. Implementing the redshift and the total mass, the correction functions return a revised cluster temperature and redshift, which sequently describe a corrected total mass. These steps are repeated until negligible changes of the properties are obtained with each iteration and the final values are adopted as best estimates.\\\indent
Through our simulations, we also investigate the deviation in the uncertainties between the results by the \textit{xspec} {\tt{error}}-command and a statistical distribution. These corrections of the uncertainties need to be considered for the data analysis of clusters with unknown redshift independent of the precision in the cluster properties, as \textit{xspec} underestimates the statistical uncertainty.\\\indent
In convolving the galaxy cluster mass function and scaling relations with the \textit{eROSITA} response, we obtain the distribution of clusters with mass and redshift as it will be observed by the instrument. Applying the scaling relations by \cite{Reichert2011}, we expect \textit{eROSITA} to detect $\sim113,400$ clusters of galaxies in total with a minimum photon number of $\eta_{\text{min}}=50$. Out of this total number of clusters, \textit{eROSITA} will provide precise temperatures with $\Delta T/\langle T_{\text{fit}}\rangle\lesssim10\%$ for $\sim1,670$ new clusters in the all-sky survey, which is equivalent to a percentage of $\sim1.5\%$ of the total amount of detected clusters. {This \textit{eROSITA} sample, consisting mainly of so far unstudied clusters, will increase the current catalogue of clusters with precise temperatures by a factor of $5-10$ depending on the refered to catalogue.\\\indent
Large samples of precise and accurate cluster data, as they will be available from the \textit{eROSITA} mission, are essential for the computation of tight constraints on the cosmological parameters. As the current simulations on the constraints which \textit{eROSITA} will implement on the cosmology do not include information on the cluster temperature yet \citep{Pillepich2011}, we aim to improve these constraints through our findings \citep[compare][]{Clerc2012} and will predict these improvements in our future work.

\begin{acknowledgements}
	We want to thank the German \textit{eROSITA} collaboration for discussions on the X-ray background model and the simulation setup, Frank Haberl and Jan Robrade for providing the instrumental response file and the \textit{eROSITA} exposure maps, respectively. Also, we want to acknowledge the discussions with Cristiano Porciani and Annalisa Pillepich on the implementation of the galaxy cluster mass function. Thanks also to Gerrit Schellenberger for the analysis of high redshift \textit{Chandra} clusters for comparison reasons. Finally, we also want to thank the referee for the detailed, constructive feedback and discussions. The presented work is funded by the \textit{Deutsche Telekom Stiftung}, by the Transregional Collaborative Research Center "The Dark Universe" \textit{TR33} and by the \textit{International Max-Planck Research School for Astronomy and Astrophysics} of Bonn and Cologne. T. H. R. ackknowledges support from the \textit{DFG} through the Heisenberg research grant RE 1462/5. L. L. ackknowledges support from the \textit{DFG} through the research grant RE 1462/6 and LO 2009/1-1. \textit{eROSITA} is funded equally by the \textit{Deutsches Zentrum f\"ur Luft- und Raumfahrt} (DLR) and the \textit{Max-Planck-Gesellschaft zur F{\"o}rderung der Wissenschaften} (MPG).
\end{acknowledgements}

\bibliographystyle{aa} 
	
\bibliography{kborm0414} 	
	
\newpage

\begin{appendix}

\section{Parameter bias}
\label{AParameterBias}
Within this section we state the estimated correction functions for the parameter bias and describe for which mass and redshift ranges these corrections apply (Tabs. \ref{TabMassRedshift} -- \ref{TabFitVarZ}). As the parameter biases are independent of the cluster mass for the simulations with unknown redshift, the correction function covers the entire simulated redshift space $-2\leqslant\log(z)\leqslant0.25$ in these cases. The functions are expressed by equation \ref{EqExpFit} with the variables $A$ and $B$ and present an approximated estimate for the bias correction.

\begin{table}[ht]
	\centering
	\begin{tabular}{c|c|c|c}
	\hline\hline
	group	 &	 mass range	 &	\multicolumn{2}{c}{redshift range in $\log(z)$}\\
			 &	in $\log(M/\text{M}_{}\odot)$	& $t_{\text{exp}}=1.6$ ks & $t_{\text{exp}}=20$ ks 	\\\hline
	1		 & 13 -- 13.4.5	 &	(-2) -- (-1.35) & (-2) -- (-0.8) \\ 
	2		 & 13.45 -- 14.05 &  (-1.7) -- (-0.8) &	(-1.7) -- (-0.2)\\
	3		 & 14.05 -- 14.65 & (-1.1) -- (-0.2) & (-1.1) -- 0.25\\
	4		 & 14.65 -- 15.25	 &  (-0.65) -- 0.25	 & (-0.65) -- 0.25\\
	5		 & 15.25 -- 15.7 &  (-0.2) -- 0.25	 & (-0.2) -- 0.25\\\hline
	\end{tabular}
	\caption{Mass and redshift ranges for the application of the individual correction functions of the parameter bias in case of known cluster redshift.}
	\label{TabMassRedshift}
\end{table}

\begin{table}[ht]
	\centering
	\begin{tabular}{c|c|c|c|c}
	\hline\hline
	group	& \multicolumn{2}{c|}{$t_{\text{exp}}=1.6$ ks} & \multicolumn{2}{c}{$t_{\text{exp}}=20$ ks} \\
			&	  $A$		&	$B$		& 	$A$		& 	$B$		\\\hline
		1	&	 50.0		&	5.25	&	2.53	&	4.34\\
		2	&	-0.05		&	2.47	&	-0.22	&	3.25\\
		3	&	-0.45		&	3.85	&	-0.41	&	2.51\\
		4	&	-0.17		&	2.02 	&	-0.22	&	2.43\\
		5	&	-0.03		&	2.56	&	-0.05	&	1.07\\\hline	
	\end{tabular}
	\caption{Parameters of the correction funtion for the simulation with known redshift.}
	\label{TabFit}
\end{table}

\begin{table}[ht]
	\centering
	\begin{tabular}{c|c|c|c|c}
	\hline\hline
	parameter	& \multicolumn{2}{c|}{$t_{\text{exp}}=1.6$ ks} & \multicolumn{2}{c}{$t_{\text{exp}}=20$ ks} \\
				&  	 $A$		&	$B$		& 	$A$		& 	$B$		\\\hline
	 temperature	&	 -0.46		&	2.29	&	-0.32	&	2.53\\
	 redshift		&	-0.28		&	2.91	&	-0.37	&	3.54\\\hline		
	\end{tabular}
	\caption{Parameters of the correction funtions for the biases in the temperature and the redshift when the cluster redshift itself is unavailable. For these simulations these biases are independent of the cluster mass.}
	\label{TabFitVarZ}
\end{table}

\section{Comparison between different scaling relations}
\label{NumberComparison}
In addition to the comparison of the number of clusters for the scaling relations by \cite{Reichert2011} and \cite{Vikhlinin2009}, we performed a thorough analysis of the distribution of galaxy clusters with mass and redshift for these two relations. For both relations a cosmology of $\Omega_{\text{m}}=0.3$, $\Omega_{\Lambda}=0,7$, $h=0.7$ and $\sigma_{\text{8}}=0.795$ is assumed. The distribution are presented for three different minimum numbers of detected photons $\eta_{\text{min}}=50$, $500$ and $1500$ (Figs. \ref{PixReichert} $\&$ \ref{PixVikhlinin}). Even though sources with as little as $50$ photon counts are assumed to be identified as galaxy clusters, a larger number of counts improves the precision and the accuracy of the reduced cluster properties. The simulation of these distributions follows the analogous setup as described in Sec. \ref{Cosmo}.\\\indent
With an increasing value for $\eta_{\text{min}}$, the total number of detected clusters declines significantly as the distribution of clusters becomes shallower and the low and intermediate mass clusters are no longer detected at the high redshifts. According to this, the total number of detected clusters decreases from $113,400$ for $\eta_{\text{min}}=50$ to $11,000$ for  $\eta_{\text{min}}=500$ and to $3,000$ for $\eta_{\text{min}}=1500$. At the same time, the maximum of the distribution shifts to lower redshift values $z<0.3$. In comparison, both scaling relations yield the same position of the maximum of the distribution, where the distribution based on the scaling relation by \cite{Reichert2011} displays a broader peak. This development results in a total number of clusters, which is $15-20 \%$ above the value for the study of the scaling relation by \cite{Vikhlinin2009} with a total number of cluster of $103,700$ for $\eta_{\text{min}}=50$, $8,900$ for $\eta_{\text{min}}=500$ and $2,300$ for $\eta_{\text{min}}=1500$.\\\indent
This analysis emphasises the strong dependence of the distribution of clusters and of the total number of detected clusters on the applied scaling relations and the defined minimum number of photons $\eta_{\text{min}}$. 
\newpage

\begin{figure}[P]
	\centering
	\vspace{-0.2cm}
	\includegraphics[scale=0.5]{figure15.pdf}\vspace{0.1cm}
	\includegraphics[scale=0.5]{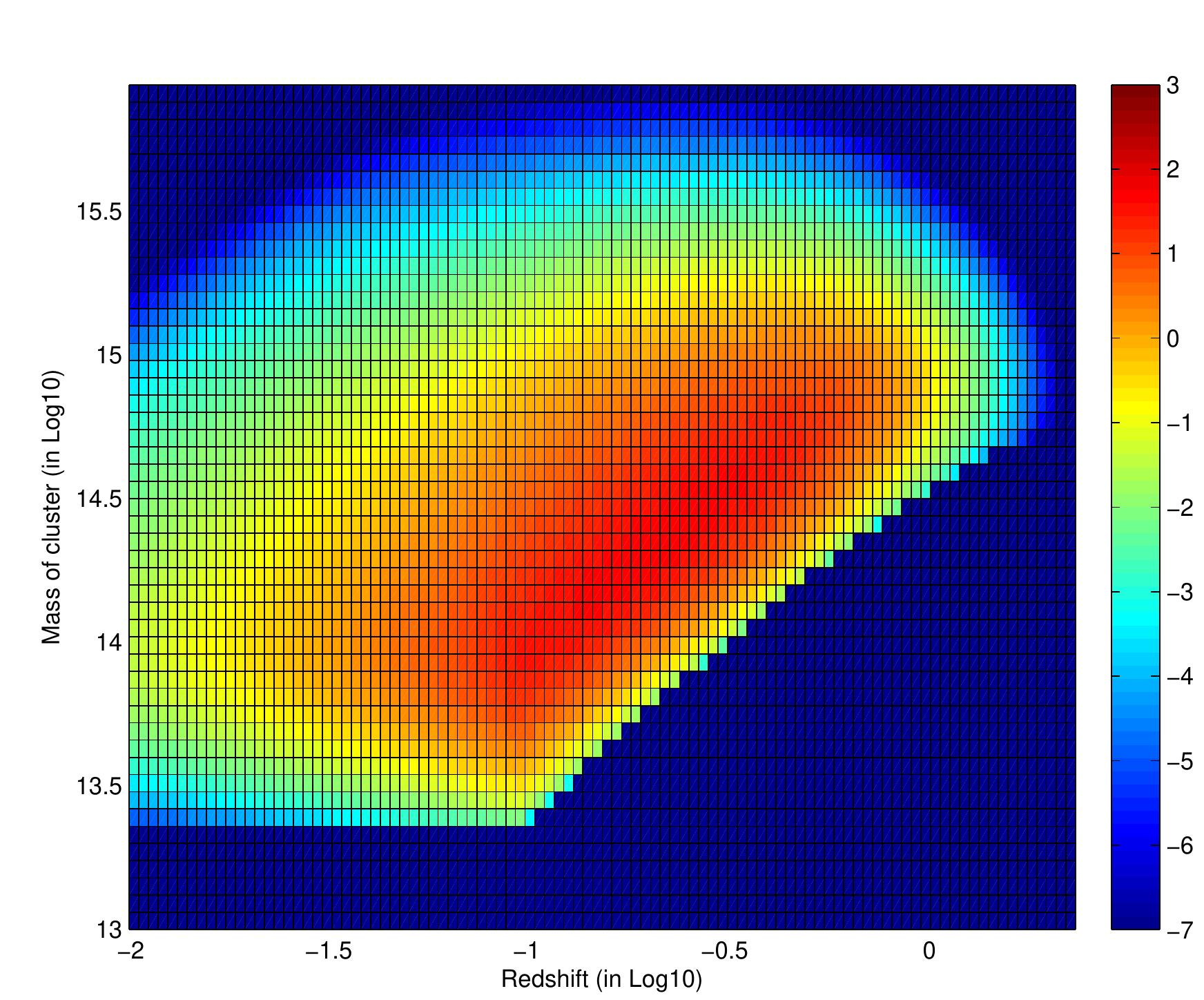}\vspace{0.1cm}
	\includegraphics[scale=0.5]{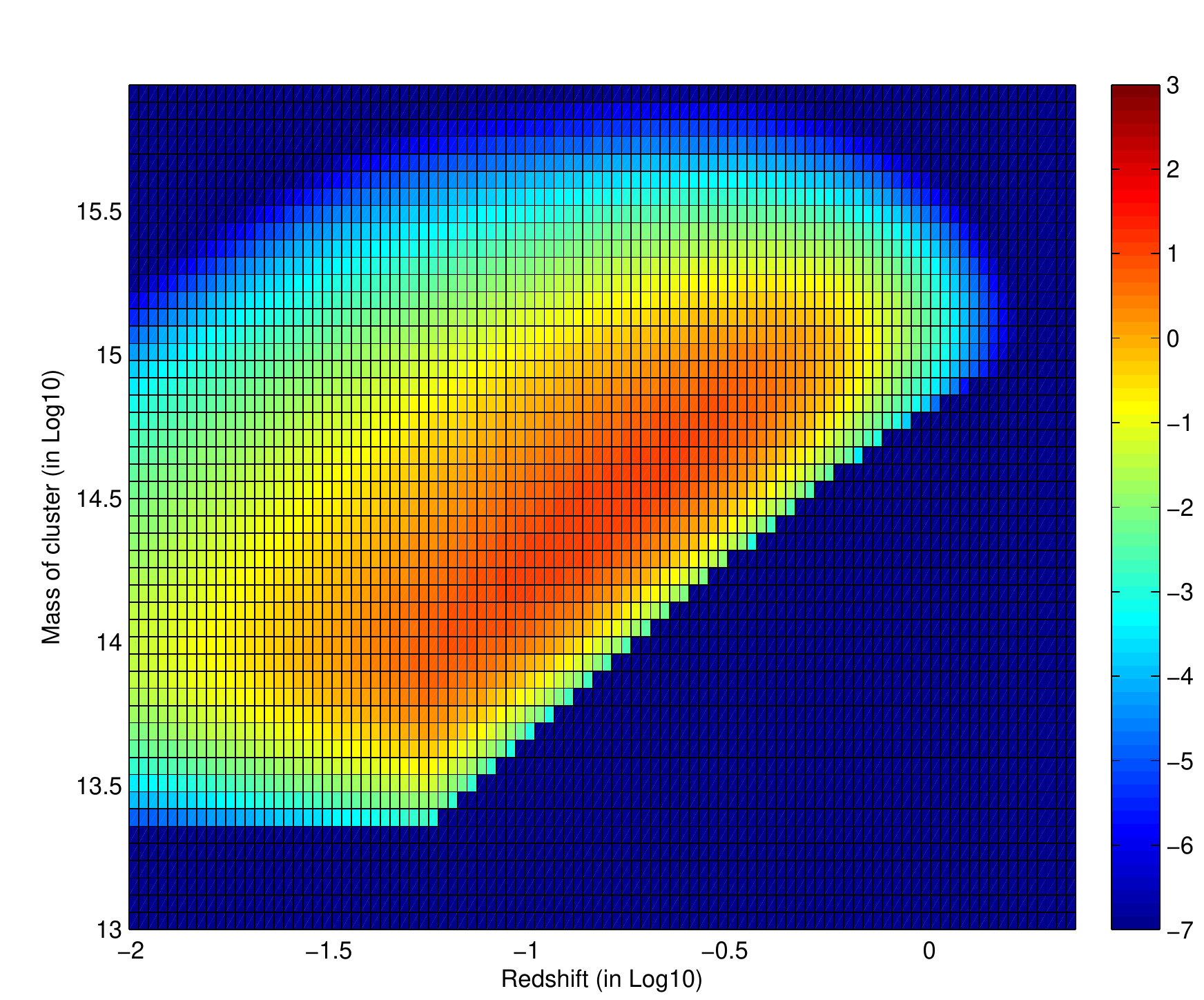}
	\caption{Distribution of galaxy clusters with mass and redshift for three different photon detection minimums $\eta_{\text{min}}=50,$  $500$ and $1500$ from top to bottom for the scaling relation by \cite{Reichert2011}. All plots are generated for a lower mass cut of $M=5\times10^{13}/h_{100}$ M$_{\odot}$ with $h_{100}=0.7$. The colour indicates the number of detected clusters in the individual bins in units of $\log_{10}$, where the cluster mass is considered in units of $\log(M/\text {M}_{\odot})$. The total number of detected clusters reads from top to bottom $N_{\text{cluster}}=113,400$ $11,000$ and $3,000$.}
	\label{PixReichert}
\end{figure}
\begin{figure}[P]
	\centering
	\vspace{0.3cm}
	\includegraphics[scale=0.5,trim=0.0cm 0cm 0.0cm 1.0cm, clip=true]{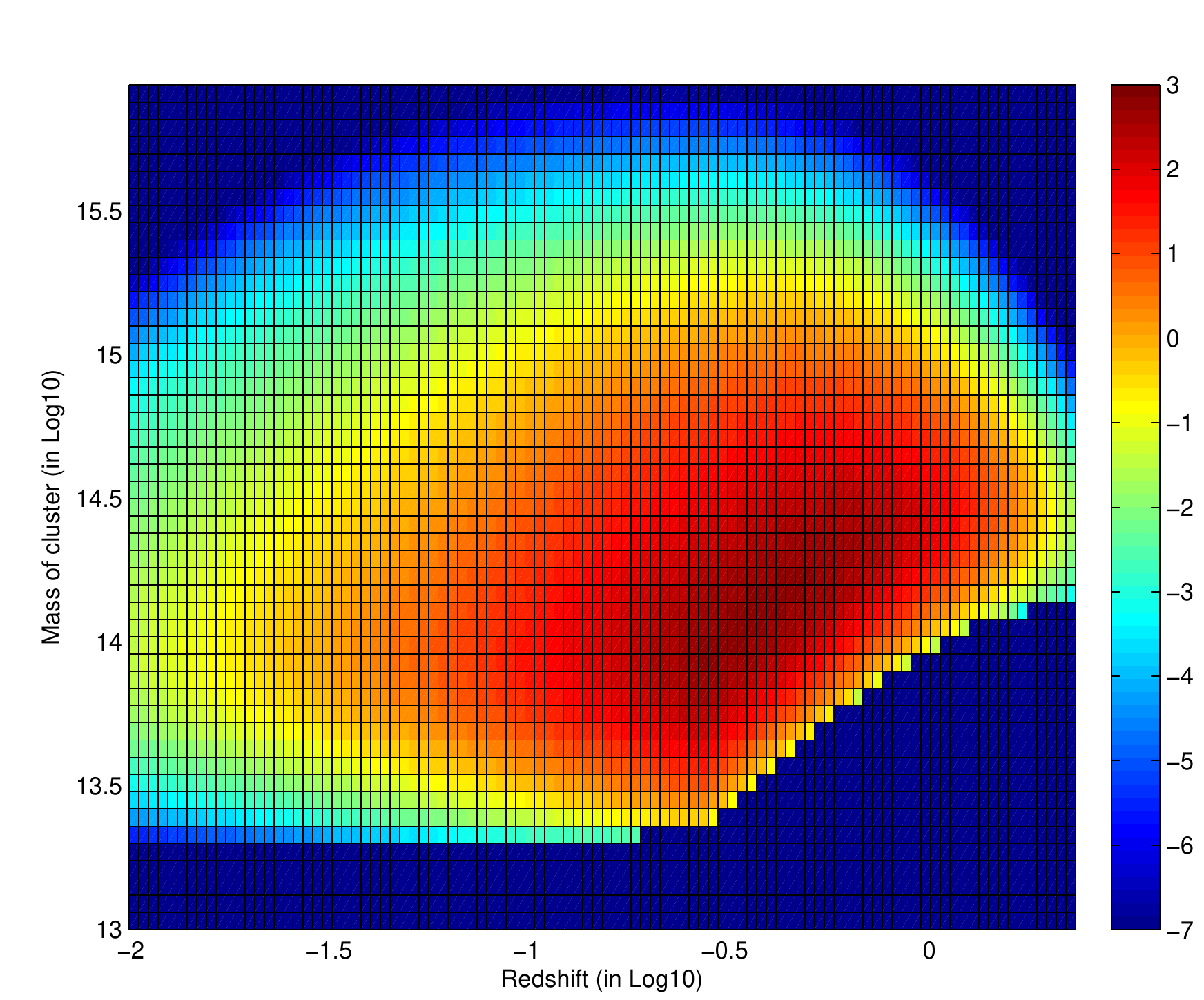}\vspace{0.7cm}
	\vspace{0.2cm}
	\includegraphics[scale=0.5,trim=0.0cm 0cm 0.0cm 1.0cm, clip=true]{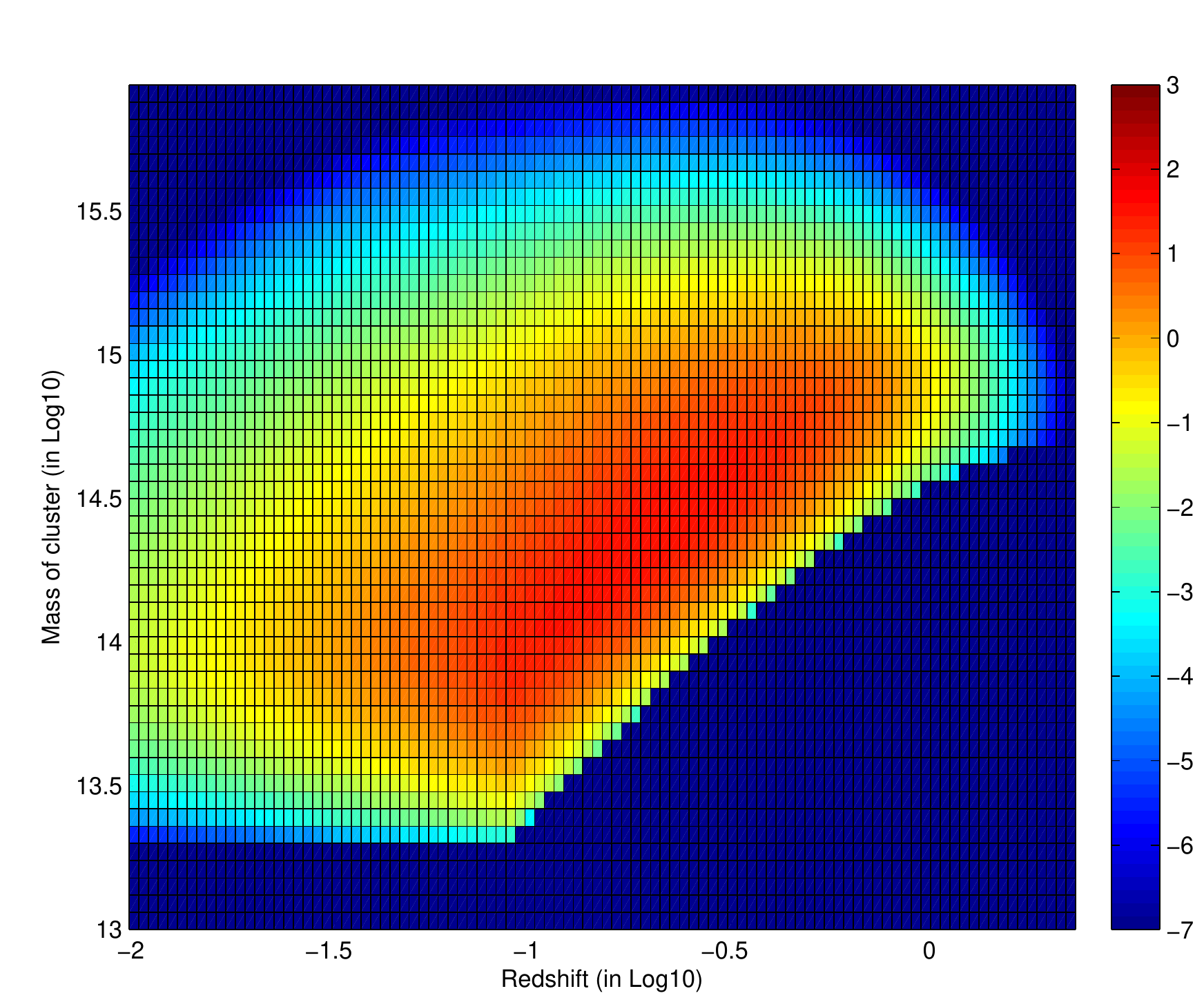}\vspace{0.3cm}
	\includegraphics[scale=0.5,trim=0.0cm 0cm 0.0cm 1.0cm, clip=true]{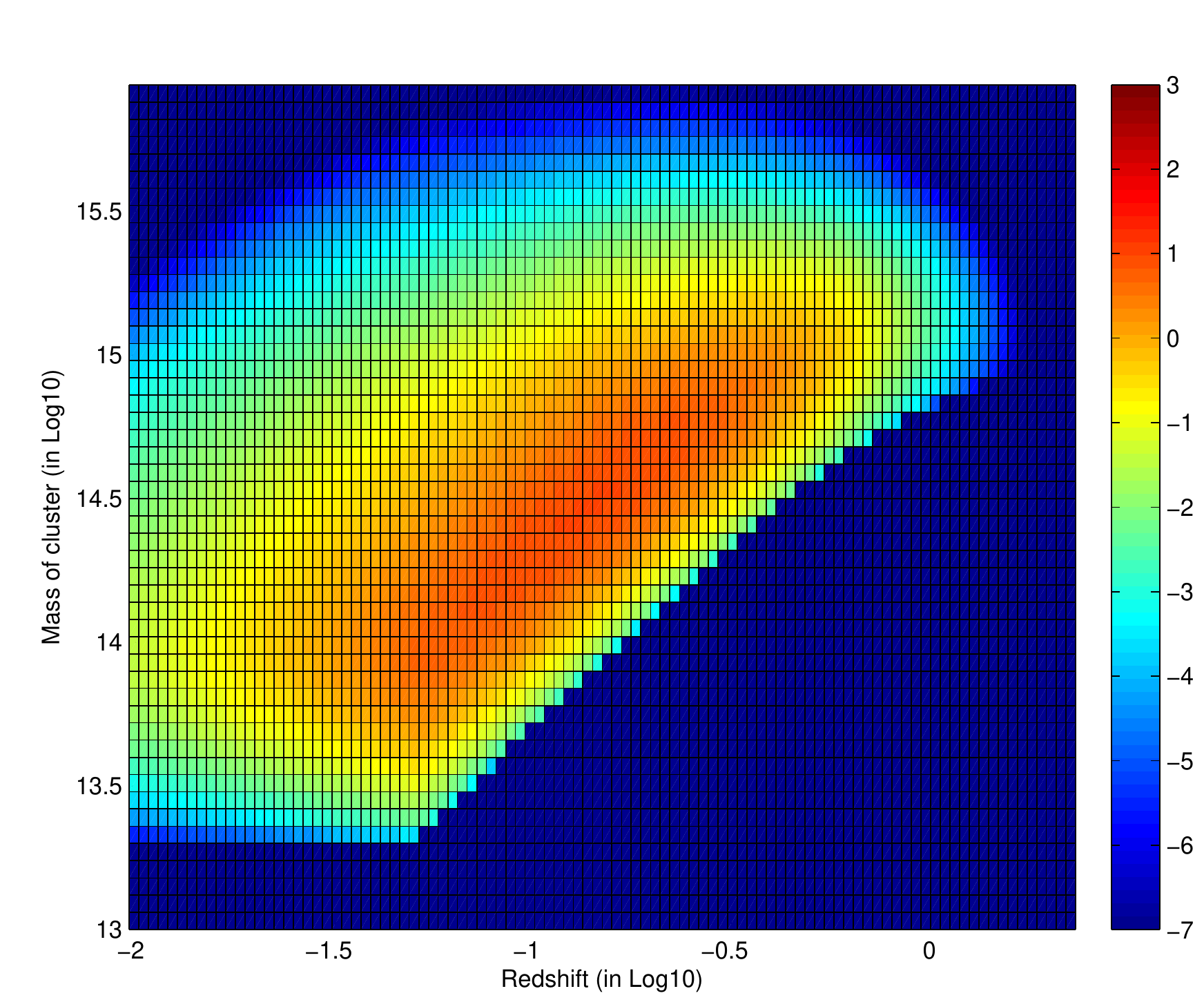}
		\caption{Distribution of galaxy clusters with mass and redshift for three different photon detection minimums $\eta_{\text{min}}=50,$  $500$ and $1500$, when applying the scaling relations by \citep{Vikhlinin2009}. All plots are generated for a lower mass cut of $M=5\times10^{13}/h_{100}$ M$_{\odot}$ with $h_{100}=0.70$, where the labeling is equivalent to Fig. \ref{PixReichert}. The total number of detected clusters reads from top to bottom $N_{\text{cluster}}=103,700$ $8,900$ and $2,300$.}
		\label{PixVikhlinin}
\end{figure}

\end{appendix}

\end{document}